\documentclass[sigconf, screen, nonacm]{acmart}

\usepackage[linesnumbered,ruled,vlined]{algorithm2e}
\usepackage{makecell}
\usepackage{enumitem}
\usepackage{tikz}
\usepackage{tikz-qtree}
\usepackage{caption}
\usepackage{subcaption}
\usepackage{tabularx}
\usepackage{pgfplots}
\pgfplotsset{compat=1.18}
\usepackage{amsmath, mathtools}
\usepackage[dvipsnames]{xcolor}
\usetikzlibrary{fit, backgrounds}
\usetikzlibrary{topaths, calc, arrows.meta, positioning, decorations.pathreplacing}

\setlist[itemize]{leftmargin=10pt}
\setlist[enumerate]{leftmargin=20pt}

\pgfplotsset{every axis/.append style={
    font=\sffamily,
    label style={font=\sffamily},
    tick label style={font=\sffamily}
}}

\usepackage{multirow}
\usepackage{graphicx}
\usepackage{float}
\usepackage{balance}
\usepackage{pgfplotstable}

\newcommand{\subject}[1]{}

\newcommand{\bnb}{bnb}

\newcommand{\nbr}{N}
\newcommand{\nnbr}{\overline{N}}
\newcommand{\Color}{\chi}

\newcommand{\pivot}{\textsf{Pivot2+}}
\newcommand{\pivottwo}{\textsf{Pivot2}}
\newcommand{\kDC}{\textsf{kDC}}
\newcommand{\kDCtwo}{\textsf{kDC2}}
\newcommand{\Ours}{\textsf{WODC}}
\newcommand{\OursS}{\textsf{WODC\textsubscript{S}}}
\newcommand{\OursE}{\textsf{WODC\textsubscript{E}}}
\newcommand{\KDClub}{\textsf{KD-Club}}
\newcommand{\MDC}{\textsf{MDC}}

\newcommand{\myparagraph}[1]{\paragraph{#1}}

\newcommand{\illegalvspace}[1]{\vspace{#1}}

\DeclareMathOperator*{\argmin}{arg\,min}

\newcommand{\redundant}[2]{\color{blue}#2 \color{black}}

\newcommand{\remove}[1]{}

\SetVlineSkip{1.8pt}

\setlist{topsep=.5ex plus .2ex minus .1ex}

\usepackage{titlesec}
\titlespacing*{\section}
  {0pt}{1.5ex plus .3ex minus .2ex}{0.75ex plus .15ex}
\titlespacing*{\subsection}
  {0pt}{1.25ex plus .3ex minus .2ex}{0.5ex plus .15ex}
\titlespacing{\subsubsection}
  {0pt}{0.5ex plus 0.2ex minus .15ex}{0.25ex plus 0.15ex}
\titlespacing{\paragraph}
  {0pt}                         %
  {0.25ex plus 0.15ex minus .15ex}  %
  {1ex}                         %

\titleformat{\subsubsection}[runin]
  {\sffamily\itshape} %
  {\thesubsubsection}               %
  {1em}                             %
  {}                                %
  [. \hspace{1em}]                   %
\titleformat{\paragraph}[runin]
  {\normalfont\itshape}      %
  {\theparagraph}            %
  {1em}                      %
  {\hspace{1em}}                         %
  [. ]                       %

\newtheoremstyle{MyStyle}%
  {1pt plus 1.5pt minus 0.5pt} %
  {1pt plus 1.5pt minus 0.5pt} %
  {\itshape} %
  {1em} %
  {} %
  {.} %
  {.5em} %
  {\textsc{\thmname{#1}\thmnumber{ #2}\thmnote{ (#3)}}} %
\theoremstyle{MyStyle}
\newtheorem{mylemma}{Lemma}[section]
\newtheorem{mytheorem}{Theorem}[section]
\newtheorem{myproperty}{Property}[section]
\newtheorem{myobservation}{Observation}[section]

\newtheoremstyle{MyStyleTwo}%
  {1pt plus 1.5pt minus 0.5pt} %
  {1pt plus 1.5pt minus 0.5pt} %
  {} %
  {1em} %
  {} %
  {.} %
  {.5em} %
  {\textsc{\thmname{#1}\thmnumber{ #2}\thmnote{ (#3)}}} %
\theoremstyle{MyStyleTwo}
\newtheorem{mydefinition}{Definition}[section]
\newtheorem{myexample}{Example}[section]

\renewenvironment{proof}[1][\proofname]{\textsc{#1. }}{\qed\vskip 1mm}

\newcommand{\OursC}{black}
\newcommand{\OurspC}{VioletRed}
\newcommand{\OurssC}{CadetBlue}
\newcommand{\OursrC}{Emerald}

\newcommand{\ParTTTTC}{Orange}
\newcommand{\ParTTTC}{OrangeRed}
\newcommand{\ParTTC}{RedViolet}
\newcommand{\ParTC}{Blue}
\newcommand{\PivotC}{red}
\newcommand{\PivottwoC}{blue}
\newcommand{\SolC}{YellowOrange}
\newcommand{\MDCC}{OrangeRed}
\newcommand{\kDCtwoC}{BlueViolet}
\newcommand{\kDCC}{OliveGreen}
\newcommand{\KDClubC}{YellowOrange}

\newcommand{\OursM}{o}
\newcommand{\OurspM}{triangle}
\newcommand{\OurssM}{triangle}
\newcommand{\OursrM}{x}
\newcommand{\ParTTTTM}{halfcircle*}
\newcommand{\ParTTTM}{halfcircle*}
\newcommand{\ParTTM}{halfcircle*}
\newcommand{\ParTM}{halfcircle*}
\newcommand{\PivotM}{pentagon}
\newcommand{\PivottwoM}{square}
\newcommand{\SolM}{-}
\newcommand{\MDCM}{pentagon}
\newcommand{\kDCtwoM}{square}
\newcommand{\kDCM}{diamond}
\newcommand{\KDClubM}{triangle}

\newcommand{\scval}{0.9}

\newcommand{\OursSc}{1.5*\scval}
\newcommand{\OurspSc}{1.5*\scval}
\newcommand{\OurssSc}{1.5*\scval}
\newcommand{\OursrSc}{1.5*\scval}
\newcommand{\ParTTTTSc}{1.3*\scval}
\newcommand{\ParTTTSc}{1.3*\scval}
\newcommand{\ParTTSc}{1.3*\scval}
\newcommand{\ParTSc}{1.3*\scval}
\newcommand{\PivotSc}{1.5*\scval}
\newcommand{\PivottwoSc}{1.2*\scval}
\newcommand{\SolSc}{1.5*\scval}
\newcommand{\MDCSc}{1.5*\scval}
\newcommand{\kDCtwoSc}{1.1*\scval}
\newcommand{\kDCSc}{1.5*\scval}
\newcommand{\KDClubSc}{1.5*\scval}

\newcommand{\OursRo}{0}
\newcommand{\OurspRo}{0}
\newcommand{\OurssRo}{180}
\newcommand{\OursrRo}{0}
\newcommand{\ParTTTTRo}{270}
\newcommand{\ParTTTRo}{180}
\newcommand{\ParTTRo}{90}
\newcommand{\ParTRo}{0}
\newcommand{\PivotRo}{0}
\newcommand{\PivottwoRo}{0}
\newcommand{\SolRo}{0}
\newcommand{\MDCRo}{180}
\newcommand{\kDCtwoRo}{45}
\newcommand{\kDCRo}{90}
\newcommand{\KDClubRo}{270}

\AtBeginDocument{%
  }

\setcopyright{acmlicensed}
\copyrightyear{2018}
\acmYear{2018}
\acmDOI{XXXXXXX.XXXXXXX}

\acmISBN{978-1-4503-XXXX-X/18/06}

\begin{document}

\title{Efficient Defective Clique Enumeration and Search with Worst-Case Optimal Search Space (Full Version)}

\author{Jihoon Jang, Yehyun Nam, and Kunsoo Park}
\authornote{Corresponding author.}
\email{{jhjang, yhnam, kpark}@theory.snu.ac.kr}
\affiliation{%
  \institution{Seoul National University}
  \city{Seoul}
  \country{Republic of Korea}
}

\author{Hyunjoon Kim}
\email{hyunjoonkim@hanyang.ac.kr}
\affiliation{%
  \institution{Hanyang University}
  \city{Seoul}
  \country{Republic of Korea}
}

\renewcommand{\shortauthors}{Jihoon Jang et al.}

\begin{abstract}
A $k$-defective clique is a relaxation of the traditional clique definition, allowing up to $k$ missing edges. This relaxation is crucial in various real-world applications such as link prediction, community detection, and social network analysis. Although the problems of enumerating maximal $k$-defective cliques and searching a maximum $k$-defective clique have been extensively studied, existing algorithms suffer from limitations such as the combinatorial explosion of small partial solutions and sub-optimal search spaces.
To address these limitations, we propose a novel clique-first branch-and-bound framework that first generates cliques and then adds missing edges. Furthermore, we introduce a new pivoting technique that achieves a search space size of $\mathcal{O}(3^{\frac{n}{3}} \cdot n^k)$, where $n$ is the number of vertices in the input graph. We prove that the worst-case number of maximal $k$-defective cliques is $\Omega(3^{\frac{n}{3}} \cdot n^k)$ when $k$ is a constant, establishing that our algorithm’s search space is \emph{worst-case optimal}.
Leveraging the diameter-two property of defective cliques, we further reduce the search space size to $\mathcal{O}(n \cdot 3^{\frac{\delta}{3}} \cdot (\delta \Delta)^k)$, where $\delta$ is the degeneracy and $\Delta$ is the maximum degree of the input graph.
We also propose an efficient framework for maximum $k$-defective clique search based on our branch-and-bound, together with practical techniques to reduce the search space.
Experiments on real-world benchmark datasets with more than 1 million edges demonstrate that each of our proposed algorithms for maximal $k$-defective clique enumeration and maximum $k$-defective clique search outperforms the respective state-of-the-art algorithms by up to four orders of magnitude in terms of processing time.
\end{abstract}

\ccsdesc[500]{Mathematics of computing~Graph algorithms}

\keywords{Defective clique, Branch-and-bound, Worst-case optimal}

\maketitle

\section{Introduction}

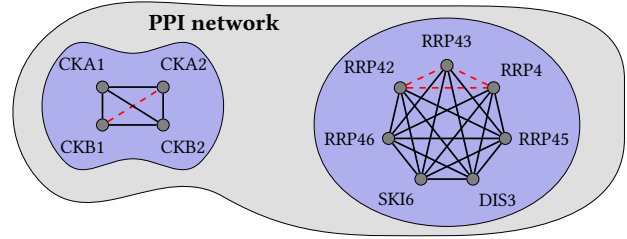
\begin{figure}
    \centering
    \scalebox{0.8}{
        \newcommand{\x}{0.7}
\newcommand{\y}{0.6}

\newcommand{\scll}{1.1}
\newcommand{\scl}{1.25}

\newcommand{\xx}{5.8}
\newcommand{\yy}{-1}

\hspace{-9mm}

\begin{tikzpicture}  
  [scale=.9,auto=center,every node/.style={circle, inner sep=0.0mm, draw, minimum width=2mm, fill=gray}, > = stealth] 
  \node [draw=none, fill=none] (u1) at (-0.8*\x, 0*\x) {};
  \node [draw=none, fill=none] (u2) at (0.8*\x, 0*\x) {};
  \node [draw=none, fill=none] (u3) at (-0.8*\x, 1*\x) {};
  \node [draw=none, fill=none] (u4) at (0.8*\x, 1*\x) {};
  \node [draw=none, fill=none] (v1) at (-0.8*\y + \xx, 0*\y + \yy) {};
  \node [draw=none, fill=none] (v2) at (0.8*\y + \xx, 0*\y + \yy) {};
  \node [draw=none, fill=none] (v3) at (-1.795984*\y + \xx, 1.252176*\y + \yy) {};
  \node [draw=none, fill=none] (v4) at (1.795984*\y + \xx, 1.252176*\y + \yy) {};
  \node [draw=none, fill=none] (v5) at (-1.436016*\y + \xx, 2.810912*\y + \yy) {};
  \node [draw=none, fill=none] (v6) at (1.436016*\y + \xx, 2.810912*\y + \yy) {};
  \node [draw=none, fill=none] (v7) at (0*\y + \xx, 3.500382*\y + \yy) {};

  \begin{scope}[fill opacity=0.5]
      \filldraw[fill=gray!50] ($(u3) + (-1*\scll, 0.8*\scll)$)
      to[out=225, in=135] ($(u1) + (-1*\scll, -1*\scll)$)
      to[out=315, in=150] ($(u2) + (1*\scll, -1.2*\scll)$)
      to[out=330, in=180] ($(\xx, \yy) + (0*\scll, -0.95*\scll)$)
      to[out=0, in=270] ($(v4) + (2*\scll, 0*\scll)$)
      to[out=90, in=270] ($(v4) + (2*\scll, 1*\scll)$)
      to[out=90, in=0] ($(v7) + (0*\scll, 1.0*\scll)$)
      to[out=180, in=0] ($(v7) + (-3.5*\scll, 1.0*\scll)$)
      to[out=180, in=45] ($(u3) + (-1*\scll, 0.8*\scll)$);
  \end{scope}
  
  \begin{scope}[fill opacity=0.5]
      \filldraw[fill=blue!50] ($(u3) + (-0.55*\scl, 0.55*\scl)$)
      to[out=225, in=135] ($(u1) + (-0.55*\scl, -0.55*\scl)$) 
      to[out=315, in=180] ($(0, -0.5*\scl)$) 
      to[out=0, in=225] ($(u2) + (0.55*\scl, -0.55*\scl)$)
      to[out=45, in=315] ($(u4) + (0.55*\scl,0.55*\scl)$)
      to[out=135, in=0] ($(0, 1*\x+0.5*\scl)$)
      to[out=180, in=45] ($(u3) + (-0.55*\scl, 0.55*\scl)$);
  \end{scope}

  \begin{scope}[fill opacity=0.5]
    \filldraw[fill=blue!50] ($(\xx, \yy) + (0, -0.7*\scl)$)
    to[out=0, in=270] ($(v4) + (1.1*\scl, 0.2*\scl)$)
    to[out=90, in=0] ($(v7) + (0, 0.7*\scl)$)
    to[out=180, in=90] ($(v3) + (-1.1*\scl, 0.2*\scl)$)
    to[out=270, in=180] ($(\xx, \yy) + (0, -0.7*\scl)$);
  \end{scope}

  \node[draw=none, fill=none] (ppi) at (1.5, 1.9) {\textbf{\Large{PPI network}}};

  \node [label={[label distance=0mm]240:CKB1}] (u1) at (-0.8*\x, 0*\x) {};
  \node [label={[label distance=0mm]300:CKB2}] (u2) at (0.8*\x, 0*\x) {};
  \node [label={[label distance=0mm]120:CKA1}] (u3) at (-0.8*\x, 1*\x) {};
  \node [label={[label distance=0mm]60:CKA2}] (u4) at (0.8*\x, 1*\x) {};

  \path[-, thick] (u1) edge (u2); 
  \path[-, thick] (u1) edge (u3); 
  \path[-, thick, dashed, red] (u1) edge (u4); 
  \path[-, thick] (u2) edge (u3); 
  \path[-, thick] (u2) edge (u4); 
  \path[-, thick] (u3) edge (u4); 

  \node [draw=none, fill=none] (x) at (0, -1.6) {};

  \node [label={[label distance=1mm]210:SKI6}] (v1) at (-0.8*\y + \xx, 0*\y + \yy) {};
  \node [label={[label distance=1mm]330:DIS3}] (v2) at (0.8*\y + \xx, 0*\y + \yy) {};
  \node [label={[label distance=1mm]180:RRP46}] (v3) at (-1.795984*\y + \xx, 1.252176*\y + \yy) {};
  \node [label={[label distance=1mm]0:RRP45}] (v4) at (1.795984*\y + \xx, 1.252176*\y + \yy) {};
  \node [label={[label distance=1mm]170:RRP42}] (v5) at (-1.436016*\y + \xx, 2.810912*\y + \yy) {};
  \node [label={[label distance=1mm]10:RRP4}] (v6) at (1.436016*\y + \xx, 2.810912*\y + \yy) {};
  \node [label={[label distance=-1.3mm]90:RRP43}] (v7) at (0*\y + \xx, 3.500382*\y + \yy) {};

  \path[-, thick] (v1) edge (v2); 
  \path[-, thick] (v1) edge (v3); 
  \path[-, thick] (v1) edge (v4); 
  \path[-, thick] (v1) edge (v5); 
  \path[-, thick] (v1) edge (v6); 
  \path[-, thick] (v1) edge (v7); 
  \path[-, thick] (v2) edge (v3); 
  \path[-, thick] (v2) edge (v4); 
  \path[-, thick] (v2) edge (v5); 
  \path[-, thick] (v2) edge (v6); 
  \path[-, thick] (v2) edge (v7); 
  \path[-, thick] (v3) edge (v4); 
  \path[-, thick] (v3) edge (v5); 
  \path[-, thick] (v3) edge (v6); 
  \path[-, thick] (v3) edge (v7); 
  \path[-, thick] (v4) edge (v5); 
  \path[-, thick] (v4) edge (v6); 
  \path[-, thick] (v4) edge (v7); 
  \path[-, thick, dashed, red] (v5) edge (v6); 
  \path[-, thick, dashed, red] (v5) edge (v7); 
  \path[-, thick, dashed, red] (v6) edge (v7);

\end{tikzpicture}  
    }
    
    \caption{A noisy protein-protein interaction (PPI) network~\cite{bader2002analyzing} containing two defective cliques as subgraphs (highlighted in blue). Missing interactions in the noisy network can be predicted by identifying defective cliques and completing the missing edges (represented by red dashed edges). After completing the missing edges, two protein complexes can be identified (left: Casein Kinase II, right: Exosome complex), each forming a clique.
    }
    \label{fig:PPI}
    \Description{}
\end{figure}

Graphs have been widely used to model relationships among entities in various domains, including social networks, biological systems, financial markets, and transportation networks. An important task in graph analysis is the identification of dense substructures, as these often reveal significant patterns or underlying relationships within the data~\cite{chang2019cohesive, lee2010survey}. For instance, dense subgraphs can correspond to densely connected communities in social networks~\cite{du2007community, bedi2016community, lancichinetti2009community}, protein complexes in biological networks~\cite{yu2006predicting, harley2001uniform}, or anomalies in financial networks~\cite{ahmed2016survey}. A clique, defined as a subgraph in which every vertex pair is connected by an edge, represents the densest possible subgraph. 
There has been growing interest in developing practical algorithms for solving clique-related problems, such as maximal clique enumeration~\cite{eppstein2010listing, deng2024accelerating, bron1973algorithm, tomita2006worst, chang2013fast, cheng2012fast}, maximum clique search~\cite{chang2019efficient, chang2020efficient, tomita2007efficient, xiang2013scalable, rossi2015parallel}, $k$-clique enumeration~\cite{yuan2022efficient, danisch2018listing, li2020ordering, wang2024efficient, nam2025dist}, and $k$-clique counting~\cite{ye2022lightning, chang2024efficient, jain2017fast, almasri2022parallel, shi2021parallel}.

In real-world applications, the definition of a clique is often overly restrictive, as real-world data is frequently noisy or incomplete~\cite{yu2006predicting}. Moreover, requiring all members of a group to be connected is overly strict, as groups with a few missing relationships can still effectively represent a community~\cite{pattillo2013clique}. To overcome this limitation, several relaxations of cliques have been proposed, including the quasi-clique~\cite{abello2002massive}, $k$-defective clique~\cite{yu2006predicting}, $k$-plex~\cite{balasundaram2011clique}, $k$-club~\cite{bourjolly2002exact}, and $k$-bundle~\cite{pattillo2013clique}.

\begin{figure}[t]
    \centering

    \newcommand{\scaleval}{0.8}
    \newcommand{\val}{0.48}

    \begin{minipage}{0.08\textwidth}
    \begin{subfigure}[b]{1\textwidth}
        \centering
        \hspace*{-2.5mm}
        \scalebox{0.8}{
            \newcommand{\x}{0.9}
\newcommand{\y}{0.75}

\begin{tikzpicture}  
  [scale=.8,auto=center,every node/.style={circle, inner sep=0.4mm, draw}, > = stealth] 
  \node [] (u1) at (0*\x, 0*\y) {$u_1$};
  \node [] (u2) at (-1*\x, -1*\y) {$u_2$};
  \node [] (u3) at (1*\x, -1*\y) {$u_3$};
  \node [] (u8) at (0*\x, -2*\y) {$u_8$};
  \node [] (u4) at (-1*\x, -3*\y) {$u_4$};
  \node [] (u5) at (1*\x, -3*\y) {$u_5$};
  \node [] (u6) at (-0.6*\x, -4.2*\y) {$u_6$};
  \node [] (u7) at (0.6*\x, -4.2*\y) {$u_7$};

  \path[-, thick] (u1) edge (u2); 
  \path[-, thick] (u1) edge (u3); 
  \path[-, thick] (u2) edge (u4); 
  \path[-, thick] (u2) edge (u8); 
  \path[-, thick] (u3) edge (u5); 
  \path[-, thick] (u3) edge (u8); 
  \path[-, thick] (u4) edge (u6); 
  \path[-, thick] (u4) edge (u7); 
  \path[-, thick] (u4) edge (u8); 
  \path[-, thick] (u5) edge (u6); 
  \path[-, thick] (u5) edge (u7); 
  \path[-, thick] (u5) edge (u8); 
  \path[-, thick] (u6) edge (u7); 
  \path[-, thick] (u6) edge (u8); 
  \path[-, thick] (u7) edge (u8); 

\end{tikzpicture}  
        }
        \caption{$G$}
        \label{fig:input-graph}
    \end{subfigure}
    \end{minipage}
    \hfill    
    \begin{minipage}[t]{0.37\textwidth}
    \begin{subfigure}[b]{1\textwidth}
        \centering
        \hspace*{-4mm}
        \scalebox{\scaleval}{
            \newcommand{\nlabel}[4]{#4}
\newcommand{\nlabelp}[5]{#5}
\newcommand{\nodefec}[1]{$u_{#1}$}
\newcommand{\defec}[1]{$\textcolor{red}{u_{#1}}$}

\begin{tikzpicture}
\tikzset{level distance=14pt, sibling distance=2pt, inner sep=1.6pt, every node/.style={rounded corners=1pt}}
\Tree 
    [ .Root 
        [ .\node[draw]{$u_{1}$}; 
            [ .\node[draw, densely dashed]{$u_{4}$}; ] 
            [ .\node[draw, densely dashed]{$u_{5}$}; ] 
            [ .\node[draw, densely dashed, fill=gray!40]{$u_{6}$}; ] 
            [ .\node[draw, densely dashed, fill=gray!40]{$u_{7}$}; ] 
            [ .\node[draw, densely dashed]{$u_{8}$}; 
                [ .\node[draw, fill=gray!40]{$u_{2}$}; ] 
            ] 
        ] 
        [ .\node[draw]{$u_{2}$}; 
            [ .\node[draw, densely dashed]{$u_{3}$}; ] 
            [ .\node[draw, densely dashed]{$u_{5}$}; ] 
            [ .\node[draw, densely dashed]{$u_{6}$}; 
                [ .\node[draw]{$u_{4}$}; 
                    [ .\node[draw, fill=gray!40]{$u_{8}$}; ] 
                ] 
            ] 
            [ .\node[draw, densely dashed]{$u_{7}$}; 
                [ .\node[draw]{$u_{4}$}; 
                    [ .\node[draw, fill=gray!40]{$u_{8}$}; ] 
                ] 
            ] 
        ] 
        [ .\node[draw]{$u_{3}$}; 
            [ .\node[draw, densely dashed]{$u_{4}$}; ] 
            [ .\node[draw, densely dashed]{$u_{6}$}; 
                [ .\node[draw]{$u_{5}$}; 
                    [ .\node[draw, fill=gray!40]{$u_{8}$}; ] 
                ] 
            ] 
            [ .\node[draw, densely dashed]{$u_{7}$}; 
                [ .\node[draw]{$u_{5}$}; 
                    [ .\node[draw, fill=gray!40]{$u_{8}$}; ] 
                ] 
            ] 
        ] 
        [ .\node[draw]{$u_{4}$}; 
            [ .\node[draw, densely dashed]{$u_{5}$}; 
                [ .\node[draw]{$u_{6}$}; 
                    [ .\node[draw]{$u_{7}$}; 
                        [ .\node[draw, fill=gray!40]{$u_{8}$}; ] 
                    ] 
                    [ .\node[draw]{$u_{8}$}; ] 
                ] 
                [ .\node[draw]{$u_{7}$}; 
                    [ .\node[draw]{$u_{8}$}; ]
                ] 
            ] 
            [ .\node[draw]{$u_{6}$};
                [ .\node[draw]{$u_{7}$}; 
                    [ .\node[draw]{$u_{8}$}; ] 
                ] 
            ]
        ] 
        [ .\node[draw]{$u_{5}$}; 
            [ .\node[draw]{$u_{6}$}; 
                [ .\node[draw]{$u_{7}$}; 
                    [ .\node[draw]{$u_{8}$}; ] 
                ] 
            ] 
        ] 
    ] 
\end{tikzpicture}
        }
        \caption{A search tree of existing algorithm~\cite{dai2023maximal} without pivoting technique.}
        \label{fig:search-tree}
    \end{subfigure}

    \illegalvspace{-1mm}

    \begin{subfigure}[b]{1\textwidth}
        \centering
        \scalebox{\scaleval}{\newcommand{\nlabel}[4]{#4}
\newcommand{\nlabelp}[5]{#5}
\newcommand{\nodefec}[1]{$u_{#1}$}
\newcommand{\defec}[1]{$\textcolor{red}{u_{#1}}$}

\begin{tikzpicture}
\tikzset{level distance=14pt, sibling distance=2pt, inner sep=1.6pt, every node/.style={rounded corners=1pt}}
\Tree 
    [ .Root 
        [ .\node[draw]{$u_{8}$}; 
            [ .\node[draw, densely dashed]{$u_{1}$}; 
                [ .\node[draw, fill=gray!40]{$u_{2}$}; ] 
            ] 
            [ .\node[draw, label={[label distance=0mm]180:$M$}]{$u_{4}$}; 
                [ .\node[draw, densely dashed, fill=gray!40]{$u_{3}$}; ] 
                [ .\node[draw, densely dashed]{$u_{5}$}; 
                    [ .\node[draw]{$u_{6}$}; 
                        [ .\node[draw, fill=gray!40]{$u_{7}$}; ] 
                    ] 
                ] 
                [ .\node[draw]{$u_{6}$}; 
                    [ .\node[draw, densely dashed, fill=gray!40]{$u_{2}$}; ] 
                    [ .\node[draw]{$u_{7}$}; ] 
                ] 
                [ .\node[draw]{$u_{2}$}; 
                    [ .\node[draw, densely dashed, fill=gray!40]{$u_{7}$}; ] 
                ] 
            ] 
            [ .\node[draw]{$u_{3}$}; 
                [ .\node[draw, densely dashed, fill=gray!40]{$u_{2}$}; ] 
                [ .\node[draw, densely dashed]{$u_{6}$}; 
                    [ .\node[draw, fill=gray!40]{$u_{5}$}; ] 
                ] 
                [ .\node[draw, densely dashed]{$u_{7}$}; 
                    [ .\node[draw, fill=gray!40]{$u_{5}$}; ] 
                ] 
            ] 
            [ .\node[draw]{$u_{5}$}; 
                [ .\node[draw, densely dashed, fill=gray!40]{$u_{2}$}; ] 
                [ .\node[draw]{$u_{6}$}; 
                    [ .\node[draw]{$u_{7}$}; ] 
                ] 
            ] 
        ] 
        [ .\node[draw]{$u_{1}$}; 
            [ .\node[draw, densely dashed]{$u_{4}$}; ] 
            [ .\node[draw, densely dashed]{$u_{5}$}; ] 
            [ .\node[draw, densely dashed, fill=gray!40]{$u_{6}$}; ] 
            [ .\node[draw, densely dashed, fill=gray!40]{$u_{7}$}; ] 
        ] 
    ] 
\end{tikzpicture}}
        \caption{A search tree of existing algorithm~\cite{dai2023maximal} with pivoting technique.}
        \label{fig:search-tree-pivot}
    \end{subfigure}
    \end{minipage}

    \begin{subfigure}[b]{0.25\textwidth}
        \hspace{-4mm}
        \scalebox{\scaleval}{\newcommand{\nlabel}[4]{#4}
\newcommand{\nlabelp}[5]{#5}
\newcommand{\nodefec}[1]{$u_{#1}$}
\newcommand{\defec}[1]{$\textcolor{red}{u_{#1}}$}

\begin{tikzpicture}
\tikzset{level distance=14pt, sibling distance=2pt, inner sep=1.6pt, every node/.style={rounded corners=1pt}}
\Tree 
    [ .Root 
        [ .\node[draw, label={[label distance=0mm]180:$M_1$}]{$u_{1}$}; 
            [ .\node[draw, label={[label distance=0mm]180:$M_2$}]{$u_{2}$}; ] 
        ] 
        [ .\node[draw]{$u_{2}$}; 
            [ .\node[draw]{$u_{4}$}; 
                [ .\node[draw, label={[label distance=0mm]180:$M_3$}]{$u_{8}$}; 
                    [ .\node[draw, densely dashed, fill=gray!40]{$u_{6}$}; ] 
                    [ .\node[draw, densely dashed, fill=gray!40]{$u_{7}$}; ] 
                ] 
            ] 
        ] 
        [ .\node[draw]{$u_{3}$}; 
            [ .\node[draw]{$u_{5}$}; 
                [ .\node[draw]{$u_{8}$}; 
                    [ .\node[draw, densely dashed, fill=gray!40]{$u_{6}$}; ] 
                    [ .\node[draw, densely dashed, fill=gray!40]{$u_{7}$}; ] 
                ] 
            ] 
        ] 
        [ .\node[draw]{$u_{4}$}; 
            [ .\node[draw]{$u_{6}$}; 
                [ .\node[draw]{$u_{7}$}; 
                    [ .\node[draw]{$u_{8}$}; 
                        [ .\node[draw, densely dashed, fill=gray!40]{$u_{5}$}; ] 
                    ] 
                    [ .\node[draw, densely dashed]{$u_{5}$}; ] 
                ] 
                [ .\node[draw]{$u_{8}$}; 
                    [ .\node[draw, densely dashed]{$u_{5}$}; ] 
                ] 
            ] 
            [ .\node[draw]{$u_{7}$}; 
                [ .\node[draw]{$u_{8}$}; 
                    [ .\node[draw, densely dashed]{$u_{5}$}; ] 
                ] 
            ] 
        ] 
        [ .\node[draw]{$u_{5}$}; 
            [ .\node[draw]{$u_{6}$}; 
                [ .\node[draw]{$u_{7}$}; 
                    [ .\node[draw]{$u_{8}$}; ] 
                ] 
            ] 
        ] 
    ] 
\end{tikzpicture}}
        \caption{A search tree of our algorithm without pivoting technique. \\}
        \label{fig:search-tree-ours}
    \end{subfigure}
    \hfill
    \begin{subfigure}[b]{0.20\textwidth}
        \scalebox{\scaleval}{\newcommand{\nlabel}[4]{#4}
\newcommand{\nlabelp}[5]{#5}

\newcommand{\nodefec}[1]{$u_{#1}$}
\newcommand{\defec}[1]{$\textcolor{red}{u_{#1}}$}

\begin{tikzpicture}
\tikzset{level distance=14pt, sibling distance=2pt, inner sep=1.6pt, every node/.style={rounded corners=1pt}}
\Tree 
    [ .Root 
        [ .\node[draw]{$u_{1}$}; 
            [ .\node[draw]{$u_{2}$}; ] 
        ] 
        [ .\node[draw, label={[label distance=0mm]180:$M_1$}]{$u_{8}$}; 
            [ .\node[draw, label={[label distance=0mm]180:$M_2$}]{$u_{2}$}; 
                [ .\node[draw, label={[label distance=0mm]180:\redundant{$M_3$}{}}]{$u_{4}$}; 
                    [ .\node[draw, densely dashed, fill=gray!40]{$u_{6}$}; ] 
                    [ .\node[draw, densely dashed, fill=gray!40]{$u_{7}$}; ] 
                ] 
            ] 
            [ .\node[draw]{$u_{3}$}; 
                [ .\node[draw]{$u_{5}$}; 
                    [ .\node[draw, densely dashed, fill=gray!40]{$u_{6}$}; ] 
                    [ .\node[draw, densely dashed, fill=gray!40]{$u_{7}$}; ] 
                ] 
            ] 
            [ .\node[draw]{$u_{7}$}; 
                [ .\node[draw]{$u_{6}$}; 
                    [ .\node[draw]{$u_{4}$}; 
                        [ .\node[draw, densely dashed, fill=gray!40]{$u_{5}$}; ] 
                    ] 
                    [ .\node[draw]{$u_{5}$}; ] 
                ] 
            ] 
        ] 
    ] 
\end{tikzpicture}}
        \caption{A search tree of our algorithm with pivoting technique.}
        \label{fig:search-tree-ours-pivot}
    \end{subfigure}

    \caption{An example graph $G$ and search trees to enumerate all maximal 1-defective cliques of size at least 4 in $G$. The label of each search tree node represents the branching vertex that is last added to the partial solution, and the vertices from the root to each node form the partial solution of that node. A node with a dashed outline indicates that adding the branching vertex introduces missing edges in its solution, and a gray-shaded node indicates that its partial solution is maximal in $G$.}
    \label{fig:search-trees}
    \Description{}
\end{figure}
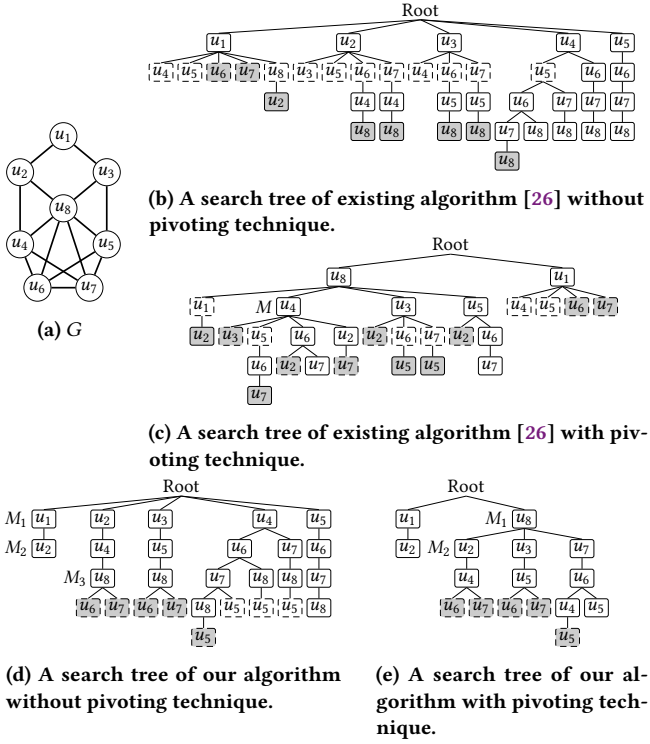

In this paper, we focus on the $k$-defective clique, which allows a subgraph to miss up to $k$ edges to be a clique. Given a graph $G=(V, E)$ with $n$ vertices and $m$ edges, a set $S$ of vertices in $G$ is a $k$-defective clique if the subgraph induced by $S$ has at least $\binom{|S|}{2} - k$ edges. For example, consider the graph $G$ in Figure~\ref{fig:input-graph}. The set $S=\{u_4, u_5, u_6, u_7, u_8\}$ of vertices is a $1$-defective clique, as the subgraph induced by $S$ has $\binom{5}{2} - 1 = 9$ edges. 
The \emph{size} of a $k$-defective clique is defined by the number of vertices it contains. 
A $k$-defective clique $S$ is called \emph{maximal} if there is no other $k$-defective clique including $S$ and is called \emph{maximum} if it has the largest size among all $k$-defective cliques; it is obvious that a maximum $k$-defective clique is also a maximal $k$-defective clique.

The analysis of real-world networks frequently involves the task of identifying defective cliques of large size. The missing edges in a defective clique can serve as a basis for predicting missing interactions or relationships within noisy networks. For instance, Yu et al.~\cite{yu2006predicting} proposed a method to predict missing interactions in protein-protein interaction (PPI) networks by completing the missing edges in defective cliques; see Figure~\ref{fig:PPI} for illustration. In addition, defective cliques have diverse applications across various domains, including community detection in social networks~\cite{gschwind2021branch, jain2020provably}, cluster identification~\cite{stozhkov2022continuous}, optimization of transportation systems~\cite{sherali2002airspace}, and statistical analysis in financial networks~\cite{boginski2005statistical, boginski2006mining}.

\linepenalty=10

The importance of finding $k$-defective cliques has motivated extensive research focused on the \emph{maximal $k$-defective clique enumeration}~\cite{dai2023maximal} and the \emph{maximum $k$-defective clique search}~\cite{gao2022exact, trukhanov2013algorithms, chang2023efficient, jin2024kd, chen2021computing, dai2024theoretically, chang2024maximum}.
The maximal $k$-defective clique enumeration problem is to enumerate all maximal $k$-defective cliques of size at least the given threshold $q$ in a graph. The maximum $k$-defective clique search problem is to find a maximum $k$-defective clique in the graph. However, both of these problems are NP-hard~\cite{trukhanov2013algorithms, yannakakis1978node}; 
as a result, solving them remains a computational bottleneck in the applications.

\myparagraph{Existing algorithms and limitations}
Although the two problems are closely related, the research on each problem has been conducted separately~\cite{gao2022exact, chang2023efficient, jin2024kd, chen2021computing, dai2023maximal, dai2024theoretically, chang2024maximum}.
The commonality between approaches for the two problems is that they find \emph{solutions} (i.e., maximum or maximal $k$-defective cliques) through a branch-and-bound algorithm~\cite{land2010automatic}. 

\linepenalty=500

A recursive call in the branch-and-bound algorithm maintains two disjoint sets of vertices: $S$ and $C$. The set $S$ represents a \emph{partial solution}, which is a $k$-defective clique. The set $C$ consists of \emph{candidate vertices}, each of which can be added to $S$, resulting in another $k$-defective clique.
The goal of the recursive call is to find all solutions that include $S$ and are included by $S \cup C$. 
For each \emph{branching vertex} $b \in C$, a recursive call is made to enumerate all solutions that include $S \cup \{b\}$.
This procedure is repeated recursively to find all solutions.

In recursive calls, the order in which branching vertices are selected is a crucial issue.
The state-of-the-art algorithms~\cite{chang2023efficient, dai2023maximal, chang2024maximum, dai2024theoretically} prioritize the selection of a branching vertex that introduces missing edges to the solution.
However, this branching rule often causes a combinatorial explosion of small partial solutions, especially in real-world networks that are sparse and contain numerous vertices.
Figure~\ref{fig:search-tree} illustrates the search tree generated by an algorithm in \cite{dai2023maximal} for enumerating all maximal 1-defective cliques of size at least 4 in the graph $G$ shown in Figure~\ref{fig:input-graph}. 
A total of 15 partial solutions of size 2 are generated, and 13 of them contain missing edges (e.g., $\{u_1, u_4\}$ and $\{u_2, u_3\}$).

\begin{table}[t]
    \setlength{\tabcolsep}{2.0pt}
    \caption{Comparison of search space size (i.e., the number of search tree nodes) generated by different algorithms, and the worst-case output size of maximal $k$-defective clique enumeration when $k$ is a constant.
    }
    \label{tab:search-space}
    \centering
    \scalebox{0.83}{
    \begin{tabular}{c|c|c|c|c|c|c|c}
    \toprule
        Algorithms & \# nodes & $k=$ & 1 & 2 & 3 & 4 & $\cdots$ \\\midrule
        \textsf{KDBB}~\cite{gao2022exact}, \KDClub{}~\cite{jin2024kd} & $\mathcal{O}(2^n)$ & -- & -- & -- & -- & -- & $\cdots$ \\
        \textsf{MADEC}~\cite{chen2021computing} & $\mathcal{O}(\alpha_k^n)$ & $\alpha_k=$ & 1.928 & 1.984 & 1.996 & 1.999 & $\cdots$ \\
        \kDC{}~\cite{chang2023efficient}, \textsf{Pivot}~\cite{dai2023maximal} & $\mathcal{O}(\beta_k^n)$ & $\beta_k=$ & 1.839 & 1.928 & 1.966 & 1.984 & $\cdots$ \\ 
        \textsf{kDC2}~\cite{chang2024maximum} & $\mathcal{O}(\beta_{k-1}^n)$ & $\beta_{k-1}=$ & 1.618 & 1.839 & 1.928 & 1.966 & $\cdots$ \\ 
        \MDC{}~\cite{dai2024theoretically} & $\mathcal{O}(\gamma_k^n)$ & $\gamma_k=$ & 1.466 & 1.755 & 1.889 & 1.948 & $\cdots$ \\ 
        Our algorithm & $\mathcal{O}(3^{\frac{n}{3}} \cdot n^k)$ & $3^{\frac{1}{3}}=$ & \textbf{1.442} & \textbf{1.442} & \textbf{1.442} & \textbf{1.442} & $\cdots$ \\ \midrule
        Worst-case output size & $\Omega(3^{\frac{n}{3}} \cdot n^k)$ & $3^{\frac{1}{3}}=$ & 1.442 & 1.442 & 1.442 & 1.442 & $\cdots$ \\ \bottomrule
    \end{tabular}
    }
\end{table}

From a theoretical perspective, existing algorithms provide upper bounds on the search space size (i.e., the number of nodes in the search tree); see Table~\ref{tab:search-space}.
However, these search spaces are sub-optimal, as their sizes do not match the worst-case output size of maximal $k$-defective clique enumeration, which we prove to be $\Omega(3^{\frac{n}{3}} \cdot n^k)$ when $k$ is a constant (Section~\ref{subsec:time-complexity}). Moreover, as $k$ increases, the search space size of existing algorithms approaches the trivial bound of $\mathcal{O}(2^n)$, which corresponds to enumerating all subsets of vertices.

\myparagraph{Contributions}
In this paper, we propose theoretically and practically efficient frameworks that address both the maximal $k$-defective clique enumeration and the maximum $k$-defective clique search.

\begin{itemize}
    \item We propose a novel branch-and-bound algorithm for maximal $k$-defective clique enumeration (Section~\ref{subsec:new-bnb}).
    A key distinction of our algorithm is that it first generates cliques and then adds vertices that introduce missing edges, whereas previous algorithms first add vertices that introduce missing edges. Figure~\ref{fig:search-tree-ours} illustrates the search tree generated by our algorithm without applying the pivoting technique. 
    All partial solutions of size no greater than $3$ are cliques (e.g., $\{u_1, u_2\}$, $\{u_2, u_4, u_8\}$, and $\{u_3, u_5, u_8\}$) and there are only six partial solutions of size $2$. In this way, our algorithm effectively mitigates the combinatorial explosion of small partial solutions. 

\linepenalty=5000

    \item We develop a pivoting technique combined with a new pivot selection strategy to reduce the search space (Section~\ref{subsec:pivot}), and we prove that the search space of our algorithm with our pivoting technique is \emph{worst-case optimal} when $k$ is a constant (Section~\ref{subsec:time-complexity}). Figure~\ref{fig:search-tree-ours-pivot} illustrates a further reduced search tree generated by our algorithm with the pivoting technique incorporated, whereas Figure~\ref{fig:search-tree-pivot} shows the search tree of \cite{dai2023maximal} with its pivoting technique.
    To establish worst-case optimality, we first prove that the worst-case number of maximal $k$-defective cliques in a graph with $n$ vertices is $\Omega(3^{\frac{n}{3}} \cdot n^k)$ when $k$ is a constant. We then show that the number of search tree nodes of our algorithm is $\mathcal{O}(3^{\frac{n}{3}} \cdot n^k)$. Since all maximal $k$-defective cliques appear as leaf nodes in our search tree, our algorithm's search space is worst-case optimal. To the best of our knowledge, this is the first work to provide a worst-case optimal search space for relaxed clique problems.

\linepenalty=500

    \item For real-world networks, which are usually sparse and have low degeneracy~\cite{eppstein2010listing}, we further reduce the search space size of our algorithm to $\mathcal{O}(n \cdot 3^{\frac{\delta}{3}} \cdot (\delta\Delta)^k)$ (Section~\ref{subsec:twohop}), where $\delta \leq \sqrt{m}$ is the \emph{degeneracy}, and $\Delta$ is the maximum degree of the input graph. This improvement uses the \emph{diameter-two property} for a $k$-defective clique of size at least $k + 2$ (Property~\ref{prop:diam2}). 

    \item We propose a novel framework to solve maximum $k$-defective clique search using our branch-and-bound algorithm for maximal $k$-defective clique enumeration (Section~\ref{sec:preprocessing}).
    To enhance the efficiency of our framework, we also incorporate two practical techniques: (1) a technique for computing an \emph{initial solution}, which is a $k$-defective clique of large size, and (2) a graph reduction technique to eliminate vertices and edges that cannot be contained in any maximum $k$-defective clique.
    Both techniques substantially reduce the search space.
\end{itemize}

We perform extensive experiments to evaluate the efficiency of the proposed algorithms and individual techniques on benchmark graph datasets (Section~\ref{sec:experiments}). The benchmark graph datasets contain ten real-world graphs with more than 1 million edges and four large-scale graphs with more than 100 million edges. 
Our algorithms, \OursE{} for maximal $k$-defective clique enumeration and \OursS{} for maximum $k$-defective clique search (where \Ours{} stands for \emph{\underline{W}orst-case \underline{O}ptimal \underline{D}efective \underline{C}lique}),  are up to four orders of magnitude faster than the state-of-the-art algorithms for the respective problem.
Additionally, we empirically demonstrate the effectiveness of our individual techniques.
\vspace{0ex plus 3ex minus 0ex}

\section{Preliminaries}\label{sec:preliminaries}

In this paper, we focus on undirected and simple graphs. 
A graph $G=(V, E)$ consists of a set $V$ of vertices and a set $E$ of edges. 
We denote by $n=|V|$ and $m=|E|$, the number of vertices and edges of $G$, respectively.
If $(u, v) \in E$, we say that $u$ is \emph{adjacent to} and a \emph{neighbor} of $v$. 
The set of neighbors of $u$ in $G$ is defined as $\nbr_G(u) = \{v \in V: (u, v) \in E\}$ and the \emph{degree} of $u$ in $G$ is $d_G(u)=|N_G(u)|$.
Given a vertex subset $S$ of $V$, the \emph{induced subgraph} $G[S]$ is the subgraph of $G$ whose vertex set is $S$ and whose edge set consists of all the edges in $E$ with both endpoints in $S$.

A vertex subset $S$ of $V$ is a \emph{clique} in $G$ if every pair of vertices is adjacent to each other (i.e., $G[S]$ has $\binom{|S|}{2}$ edges). In this paper, we focus on a $k$-defective clique, which was first proposed in \cite{yu2006predicting} and has been receiving significant attention recently~\cite{chen2021computing, gao2022exact, dai2023maximal, chang2023efficient, jin2024kd, dai2024theoretically, chang2024maximum}.

\begin{mydefinition}[$k$-defective clique]
    Given a graph $G=(V, E)$ and an integer $k$, a vertex subset $S$ of $V$ is a \emph{$k$-defective clique} in $G$ if $G[S]$ misses at most $k$ edges (i.e., $G[S]$ has at least $\binom{|S|}{2} - k$ edges).
\end{mydefinition}

\redundant{Clearly, $0$-defective cliques are simply cliques. A $k$-defective clique $S$ in $G$ is \emph{maximal} if there does not exist any proper superset $S' \subseteq V$ of $S$ (i.e., $S \subset S'$) such that $S'$ is a $k$-defective clique in $G$.}{}

We now present two properties of the $k$-defective clique, which are essential for explaining our algorithm. 

\begin{myproperty}[Hereditary]\label{prop:hereditary}
    For any $k$-defective clique $S$ in $G$, every subset $H$ of $S$ is also a $k$-defective clique in $G$.
\end{myproperty}

By Property~\ref{prop:hereditary}, we can verify the maximality of a $k$-defective clique $S$ by checking that there is no vertex $u \in V \setminus S$ such that $S \cup \{u\}$ is also a $k$-defective clique in $G$. \remove{For example, the $1$-defective clique $S=\{u_4, u_5, u_6, u_7, u_8\}$ in Figure~\ref{fig:input-graph} is maximal because adding any vertex from $V \setminus S = \{u_1, u_2, u_3\}$ to $S$ does not result in a $1$-defective clique.}

\begin{myproperty}[Diameter-two property~\cite{chen2021computing}]\label{prop:diam2}
    For any $k$-defective clique $S$ in $G$ with $|S| \ge k + 2$, its diameter is at most two (i.e., any two non-adjacent vertices must have a common neighbor in $S$).
\end{myproperty}

By Property~\ref{prop:diam2}, any $k$-defective clique of size greater than or equal to $k+2$ must be both connected and dense. It is worth noting that in real-world applications, we are often interested in identifying $k$-defective cliques of large size that are connected and dense~\cite{dai2023maximal, conte2018d2k}. Thus, it is natural to focus on finding such $k$-defective cliques.

To facilitate the presentation, we introduce additional notations. 
We denote the set of common neighbors of a vertex subset $S \subseteq V$ in $G$ by $N_G(S) = \bigcap_{u \in S}N_G(u)$. 
We say that a graph $\overline{G} = (V, \overline{E})$ is a \emph{complement graph} of $G = (V, E)$ if $\overline{E} = \{(u, v) \in V \times V: (u, v) \notin E \text{ and $u \neq v$}\}$. If $(u, v) \in \overline{E}$, we say that $u$ is a \emph{non-neighbor} of $v$. The set of non-neighbors of $u$ in $G$ is denoted by $\nnbr_G(u) = \{v \in V: (u, v) \in \overline{E}\}$.
We denote the set of edges in the induced subgraph $\overline{G}[S]$ by $\overline{E}_G(S)$. 
For brevity, we omit the subscript $G$ from the notations when the context is clear. 
For any vertex subsets $S$ and $C$ of $V$ and a vertex $u$ in $V$, 
we denote by $N_C(u)$ the set of $u$'s neighbors in $C$, by $N_C(S)$ the set of common neighbors of $S$ that are in $C$, and by $\nnbr_C(u)$ the set of $u$'s non-neighbors in $C$.

\myparagraph{Degeneracy ordering and $s$-core} We describe the concepts of degeneracy ordering and the $s$-core.
\begin{mydefinition}[Degeneracy ordering~\cite{matula1983smallest}]\label{def:degen-order}
    Given a graph $G=(V, E)$, an ordering $(v_1, v_2, \ldots, v_n)$ of its vertices $V$ is a \emph{degeneracy ordering} if, for each $1 \le i \le n$, $v_i$ is a vertex with the smallest degree in the subgraph of $G$ induced by vertices $\{v_i, v_{i + 1}, \ldots, v_n\}$.
\end{mydefinition}

\begin{mydefinition}[$s$-core~\cite{seidman1983network}]
Given a graph $G=(V, E)$ and an integer $s$, the \emph{$s$-core} of $G$ is the maximal subgraph $g$ of $G$ in which every vertex in $g$ has degree $d_{g}(u) \ge s$.
\end{mydefinition}

The degeneracy ordering and $s$-cores can be computed in $\mathcal{O}(m)$ time~\cite{matula1983smallest, chang2023efficient} by iteratively removing a vertex with the smallest degree from $G$ and appending it to the end of the ordering. The largest $s$ such that $G$ includes a non-empty $s$-core is known as the \emph{degeneracy} of $G$, denoted $\delta$. \remove{For the graph $G$ in Figure~\ref{fig:input-graph}, $(u_1, u_2, u_3, u_4, u_5, u_6, u_7, u_8)$ is a degeneracy ordering of $G$. Additionally, $G$ is the $2$-core, and the subgraph of $G$ induced by $\{u_4, u_5, u_6, u_7, u_8\}$ is the 3-core of $G$.} \remove{For the graph $G$ in Figure~\ref{fig:input-graph}, the degeneracy $\delta$ of $G$ is $3$ since the $4$-core of $G$ is empty.}

Given a degeneracy ordering $(v_1, v_2, \ldots, v_n)$ of $G$, we say that $v_j$ is a \emph{forward neighbor} of $v_i$ if $v_j$ is a neighbor of $v_i$ and $i < j$. We denote the set of forward neighbors of $v_i$ by $N^+_G(v_i) = N_G(v_i) \cap \{v_{i + 1}, v_{i + 2}, \ldots, v_n\}$. The number of forward neighbors is upper bounded by the degeneracy $\delta$ of $G$, i.e., $|N^+(v_i)| \le \delta$ for every $1 \le i \le n$~\cite{chang2019cohesive}. \remove{In the graph $G$ shown in Figure~\ref{fig:input-graph}, $u_4$ has three forward neighbors, i.e., $N^+(u_4) = \{u_6, u_7, u_8\}$.}

\myparagraph{Graph coloring} 
Given a graph $G = (V, E)$, a mapping $\Color: V \rightarrow \mathbb{N}$ is called a \emph{coloring} of $G$ if $\Color(u) \neq \Color(v)$ for every edge $(u, v) \in E$. The number of distinct colors used by $\Color$ is denoted by $|\Color|$.

Given a degeneracy ordering, one way to compute a coloring of a graph is to assign colors greedily to the vertices in reverse order, assigning each vertex the smallest available value~\cite{chang2019efficient}.
Figure~\ref{fig:input-graph-preprocessing} shows an example graph with a degeneracy ordering $(u_1, u_2, \ldots, u_9)$ and its coloring $\Color$ obtained by applying this method.
In this way, we can obtain a coloring where the number of distinct colors $|\Color|$ is upper bounded by $\delta + 1$, where $\delta$ is the degeneracy of $G$.

Frequently used notations are summarized in Table~\ref{tab:notations}.

\begin{table}
    \setlength{\tabcolsep}{3.5pt}
    \caption{Frequently used notations.}
    \centering
    \begin{tabular}{cl}
    \hline
        Notation & Meaning \\ \hline
        $G=(V, E)$ & input graph with vertex set $V$ and edge set $E$ \\ 
        $n, m$ & number of vertices and edges in $G$ \\ 
        $\delta, \Delta$ & degeneracy and maximum degree of $G$ \\ 
        $S \subseteq V$ & (partial) solution, which is $k$-defective clique \\ 
        $C, X \subseteq V \setminus S$ & sets of candidate vertices \\
        $N_C(u)$ & set of $u$'s neighbors that are in $C$ \\
        $N_C(S)$ & set of common neighbors of $S$ that are in $C$ \\
        $\nnbr_C(u)$ & set of $u$'s non-neighbors that are in $C$ \\
        $\nnbr_C[u]$ & $\nnbr_C(u) \cup \{u\}$ \\
        $\overline{E}(S)$ & set of edges in the subgraph of $\overline{G}$ induced by $S$
        \\\hline
    \end{tabular}
    \label{tab:notations}
\end{table}

\subsection{Problem Statements}
In this paper, we address two fundamental problems in computing $k$-defective cliques of large size in a given graph.

\noindent
\textbf{Maximal $k$-Defective Clique Enumeration.} Given a graph $G=(V, E)$ and positive integers $k$ and $q$ (with $q \ge k + 2$), the maximal $k$-defective clique enumeration problem is to enumerate all maximal $k$-defective cliques of size at least $q$.

\noindent
\textbf{Maximum $k$-Defective Clique Search.} Given a graph $G=(V, E)$ and a positive integer $k$, the maximum $k$-defective clique search problem is to compute a largest $k$-defective clique in $G$. 

\subsection{Related Works}\label{subsec:related-works}

\myparagraph{Maximal (maximum) clique enumeration (search)}
The maximal clique enumeration problem has been extensively studied in recent decades. The most notable algorithm for this problem is the Bron-Kerbosch (BK) algorithm~\cite{bron1973algorithm}, a seminal backtracking algorithm that incorporates a pivoting technique to reduce the search space. Tomita et al.~\cite{tomita2006worst} improved upon this by introducing a carefully designed pivot selection strategy, achieving a worst-case optimal time complexity of $\mathcal{O}(3^{\frac{n}{3}})$, as there exists a graph with $\Theta(3^{\frac{n}{3}})$ maximal cliques. Eppstein et al.~\cite{eppstein2010listing} proposed a variant of BK algorithm based on degeneracy ordering, achieving a time complexity of $\mathcal{O}(n \cdot 3^{\frac{\delta}{3}})$. In addition to these theoretical advancements, various optimization techniques have been developed~\cite{deng2024accelerating, wang2025maximal, naude2016refined, cheng2012fast, das2018shared, lessley2017maximal, san2018efficiently}. 
The maximum clique search problem also has been extensively studied both from a theoretical perspective~\cite{tarjan1977finding, jian19862, robson1986algorithms, robson2001finding} and a practical perspective~\cite{chang2019efficient, li2017minimization, rossi2015parallel, san2016new, tomita2017efficient, lu2017finding}.
However, these algorithms cannot be directly applied to our defective clique problems. 

\linepenalty=10

\myparagraph{Maximal (maximum) relaxed clique enumeration (search)}
In real-world applications, the definition of a clique is often too restrictive. To address this, various relaxations have been proposed, including quasi-cliques~\cite{abello2002massive}, $k$-plexes~\cite{balasundaram2011clique}, $k$-clubs~\cite{bourjolly2002exact}, and $k$-bundles~\cite{pattillo2013clique}. To solve these problems, numerous practical algorithms have been developed~\cite{chang2022efficient, gao2024maximum, conte2017fast, wang2022listing, zhou2021improving, dai2022scaling, conte2018d2k}. 
Some of these problems can be formulated as integer linear programs (ILP)~\cite{veremyev2016exact, sherali2006polyhedral, stozhkov2022continuous}. However, ILP-based approaches are known to be less scalable than branch-and-bound methods, as demonstrated in \cite{chen2021computing}.
Recently, branch-and-bound algorithms with search space sizes smaller than the trivial $\mathcal{O}(2^n)$ bound have been proposed for each relaxation~\cite{xiao2017fast, dai2023maximal, liu2025efficient, wang2023fast, yu2023fast, zhou2020enumerating, chang2024maximum}. However, %
none of these algorithms have been proven to be worst-case optimal.

\paragraph{Graph mining systems}
\linepenalty=10
Numerous graph mining systems have been developed to efficiently support subgraph mining tasks,
\linebreak 
including the discovery of relaxed cliques such as quasi-cliques and defective cliques.
\linepenalty=500

Arabesque~\cite{teixeira2015arabesque} is a distributed graph mining platform that automates large-scale subgraph exploration through a high-level API, supporting tasks such as frequent subgraph mining, motif counting, and clique enumeration. Peregrine~\cite{jamshidi2020peregrine} introduces a pattern-aware exploration paradigm, which leverages the semantics of user-defined patterns to guide the exploration process and minimize unnecessary computations.

\linepenalty=5000
T-thinker~\cite{khalil2022parallel} and Contigra~\cite{che2024contigra} are graph mining frameworks that support quasi-clique mining and can be extended to other relaxed clique models. T-thinker employs pruning strategies specifically tailored for quasi-cliques. Contigra is designed to handle pattern mining (including quasi-cliques) with containment constraints such as maximality. It introduces search space pruning techniques to eliminate redundant computations that arise during maximality checking, and properties such as hereditary can also be leveraged within the exploration strategy to further enhance efficiency.
\linepenalty=500

\linepenalty=500

\section{A New Branch-and-Bound Algorithm}
In this section, we present our novel branch-and-bound algorithm for solving maximal $k$-defective clique enumeration.

\subsection{A Novel Clique-First Approach}\label{subsec:new-bnb}
We first present our new approach which first generates cliques and then adds missing edges to them. 
The main observation underlying our clique-first approach is as follows. 

\begin{myobservation}
A $k$-defective clique of size $q$ must contain a clique of size $q - k$.
\end{myobservation}

\begin{proof}
    Let $S$ be a $k$-defective clique of size $q$. By definition, $S$ contains at most $k$ missing edges.
    Consider the set of vertices incident to these missing edges. By removing one endpoint from each missing edge, we remove at most $k$ vertices from $S$, and the remaining vertices form a clique of size at least $q - k$.
    Therefore, a $k$-defective clique of size $q$ contains a clique of size $q - k$.
\end{proof}

To briefly illustrate the idea of our approach, consider a graph with $n=100$ vertices and $k=3, q=10$.
Then any $3$-defective clique of size $q=10$ must have a clique of size 7. The remaining task is to choose 3 vertices (from 93 vertices) with at most 3 missing edges, which has a much smaller search space than choosing 10 vertices from 100 vertices with at most 3 missing edges.

When combined with clique listing, which can be done much faster (also has much smaller search space) than defective clique listing, our clique-first approach leads to a substantially smaller search space.

We now formally introduce our branch-and-bound algorithm.
We refer to an output (i.e., a maximal $k$-defective clique) of the problem as a \emph{solution} and a vertex subset of a solution which is a $k$-defective clique and possibly non-maximal as a \emph{partial solution}. Given a partial solution $S$, we say that a vertex $u \in V \setminus S$ is a \emph{candidate vertex} if $S \cup \{u\}$ is also a $k$-defective clique.

\begin{algorithm}[t]
    \caption{Our branch-and-bound algorithm}
    \label{alg:backtrack}
    \SetKwProg{myproc}{Procedure}{}{}
\SetKwFunction{clique}{\bnb{}}
\SetKwFunction{reduce}{reduce}
\SetKwInOut{KwIn}{Input}
\SetKwInOut{KwOut}{Output}
\SetKw{Continue}{continue}
\SetKw{output}{output}
\SetKw{report}{report}
\SetKwFunction{update}{refine}
\SetKwFunction{upperbound}{upper-bound}
\KwIn{A graph $G=(V, E)$ and two integers $k, q$}
\KwOut{Every maximal $k$-defective clique in $G$ of size $\ge q$}
\clique{$\emptyset, V, \emptyset$}\;
\myproc{\clique{$S, C, X$}}{
\lIf{$C \cup X = \emptyset$ and $|S| \ge q$}{\output{$S$}}\label{line:report}
If $N_C(S) = \emptyset$, set $B \gets C$. Otherwise, select a pivot vertex $p$ from $N_C(S)$ with the fewest non-neighbors in $N_C(S)$, then set $B \gets \nnbr_C[p]$\;\label{line:pivot}
\While{$B \neq \emptyset$\label{line:while}}{
    \lIf{\textbf{UB1} of $(S, C)$ is less than $q$}{\Return{}}\label{line:ub2}
    $b \gets$ a vertex in $B$ adjacent to all vertices in $S$. If there is no such a vertex, then choose an arbitrary vertex from $B$\;\label{line:branching}
    \tcp{a vertex $b$ that introduces a missing edge is selected only when $N_B(S) = \emptyset$.}
    
    $C', X' \gets$ \update{$S, C, X, b$}\;\label{line:refine}
    \clique{$S \cup \{b\}, C', X'$}\; \label{line:recursive}
    $B\gets B \setminus \{b\}$; $C \gets C \setminus \{b\}$; $X \gets X \cup \{b\}$\; \label{line:moving}
}
}
\myproc{\update{$S, C, X, b$}}{
    $C' \gets \{u \in C \setminus \{b\}: S \cup \{b, u\} \text{ is a $k$-defective clique}\}$\;
    $X' \gets \{u \in X: S \cup \{b, u\} \text{ is a $k$-defective clique}\}$\;
    \Return{$C', X'$}\;
}

\end{algorithm}

Algorithm~\ref{alg:backtrack} shows the pseudocode of our branch-and-bound algorithm.
A recursive call to our branch-and-bound algorithm takes as input a tuple of three disjoint sets of vertices $(S, C, X)$, where $S \subseteq V$ is a partial solution, which is a $k$-defective clique, and $C \cup X = \{u \in V \setminus S: S \cup \{u\} \text{ is a $k$-defective clique}\}$ is the set of candidate vertices. We say that the tuple $(S, C, X)$ is an \emph{instance} of the recursive call.
The candidate vertices in $C$ have not yet been explored, while those in $X$ have already been explored in previous recursions. That is, the vertices in $X$ must not be added to $S$ in its recursive calls to prevent the duplicate output of the same $k$-defective cliques.
The goal of the recursive call $(S, C, X)$ is to output every maximal $k$-defective clique that includes $S$ and is included by $S \cup C$.

For each recursive call $(S, C, X)$, the algorithm checks the maximality of $S$ by verifying whether $C \cup X = \emptyset$ and, according to Property~\ref{prop:hereditary}, outputs $S$ as a solution if this condition holds and $|S| \ge q$ (Line~\ref{line:report}). 
Then, a set of branching vertices $B \subseteq C$ is computed (Line~\ref{line:pivot}); for now, suppose that $B$ is set to $C$, i.e., the pivoting technique (details will be provided in Section~\ref{subsec:pivot}) is not applied.
For each branching vertex $b \in B$, the recursive call $(S \cup \{b\}, C', X')$ is made, where $C' \subseteq C \setminus \{b\}$, $X' \subseteq X$, and $C' \cup X' = \{u \in V \setminus (S \cup \{b\}): S \cup \{b, u\} \text{ is a $k$-defective clique}\}$ is the refined set of candidate vertices. The recursive call $(S \cup \{b\}, C', X')$ enumerates every maximal $k$-defective clique that includes $S \cup \{b\}$ and is included by $S \cup \{b\} \cup C'$.
After the recursive call, $b$ is removed from $B$ and moved from $C$ to $X$ (Lines~\ref{line:branching}--\ref{line:moving}).

Since we aim to find $k$-defective cliques of size at least $q$, we compute an upper bound on the size of a $k$-defective clique that includes $S$ and is included by $S \cup C$ (Line~\ref{line:ub2}). 
Whenever this upper bound is less than $q$, the remaining recursive calls are pruned, as they cannot generate any solution of size at least $q$.
This upper-bound-based pruning is applied to $(S, C)$ in Line~\ref{line:ub2}, in order to reduce the search space efficiently.

Our algorithm uses the graph coloring-based upper bounding technique \textbf{UB1} proposed by Lijun Chang~\cite{chang2023efficient}. Given a tuple $(S, C)$, \textbf{UB1} returns $|S| + c + r$, where $c \le |N_C(S)|$ and $r \le k - |\overline{E}(S)|$ are both non-negative integers.

In Line~\ref{line:branching}, our algorithm selects a branching vertex from $B$ that is adjacent to all vertices in $S$, if such a vertex exists. That is, our algorithm first branches on vertices that do not introduce missing edges to the partial solution. A missing edge is introduced to the partial solution only after all such vertices have been processed.
Thus, our algorithm first generates cliques and then adds missing edges.

\begin{myexample}\label{ex:nopivot}
    We are given the graph $G$ in Figure~\ref{fig:input-graph}, with $k=1$ and $q=4$. Figure~\ref{fig:search-tree-ours} shows the search tree of Algorithm~\ref{alg:backtrack} without the pivoting technique. 
    At the root node, the first recursive call, $M_1 = (\{u_1\}, \{u_2, u_3, \ldots, u_8\}, \emptyset)$, is made. The node $M_1$ first branches with the branching vertex $u_2$, as it is adjacent to the vertex $u_1$ in $S$. 
    This generates the next recursive call $M_2 = (\{u_1, u_2\}, \{u_3, u_4, u_8\}, \emptyset)$. 
    However, the recursive calls of $M_2$ are pruned since the upper bound of $(\{u_1, u_2\}, \{u_3, u_4, u_8\})$ is $|S| + c + r = 2 + 0 + 1 = 3$, which is less than $q=4$.
    After backtracking to $M_1$, $u_2$ is moved from $C$ to $X$, and the upper bound of $(\{u_1\}, \{u_3, u_4, \ldots, u_8\})$ is computed in Line~\ref{line:ub2}. Since the upper bound is $3$, the remaining recursive calls of $M_1$ are pruned.
    After backtracking to the root node, a second recursive call, $(\{u_2\}, \{u_3, u_4, \ldots, u_8\}, \{u_1\})$, is generated, and the process continues down to node $M_3 = (\{u_2, u_4, u_8\}, \{u_6, u_7\}, \emptyset)$. In $M_3$, two recursive calls, $(\{u_2, u_4, u_8, u_6\}, \emptyset, \emptyset)$ and $(\{u_2, u_4, u_8, u_7\}, \emptyset, \emptyset)$, are made, and two solutions are output in Line~\ref{line:report}. 
    In this way, the search tree outputs five solutions (gray-shaded).
    In this search tree, all partial solutions of size at most $3$ are cliques, as any partial solution of size at most $3$ that contains missing edges is pruned in Line~\ref{line:ub2}.
\end{myexample}

\myparagraph{Efficient Implementation of \texttt{refine}}
Our implementation of \texttt{refine} in Algorithm~\ref{alg:backtrack} follows the implementation of \cite{dai2023maximal}, which requires a set intersection operation between $C \cup X \setminus \{b\}$ and the neighbors of $b$. The key difference is that our algorithm first checks a simple condition before refinement to determine whether both $C$ and $X$ are empty.
Let $w$ be the minimum number of non-neighbors in $S$ among all vertices in $C \cup X \setminus \{b\}$, i.e., $w = \min_{u \in C \cup X \setminus \{b\}} |\nnbr_S(u)|$. Then, for any vertex $u \in C \cup X \setminus \{b\}$, $S \cup \{b, u\}$ contains at least $w + |\nnbr_S(b)| + |\overline{E}(S)|$ missing edges. We check whether $w + |\nnbr_S(b)| + |\overline{E}(S)| > k$ holds; if so, no vertex in $C \cup X \setminus \{b\}$ can be a candidate vertex for $S \cup \{b\}$. In such cases, we set both $C'$ and $X'$ to the empty set, thereby eliminating unnecessary set intersection operations.

We note that there is still room for improvement in the search space of Algorithm~\ref{alg:backtrack}: in the search tree of Figure~\ref{fig:search-tree-ours}, nearly all subsets of the 1-defective clique $\{u_4, u_5, u_6, u_7, u_8\}$ are generated, but most are non-maximal and therefore unnecessary. 
To address this issue, we introduce a new pivoting technique for defective cliques that significantly reduces the search space.

\subsection{A New Pivoting Technique for Defective Cliques}\label{subsec:pivot}

\subject{Explain the classical pivoting technique} 
\subject{Challenges for the worst-case optimal algorithms for relaxed clique problem}
A technique called \emph{pivoting} was originally developed to reduce the search space of the BK algorithm~\cite{bron1973algorithm} to enumerate all maximal cliques.
Tomita et al.~\cite{tomita2006worst} proved that the BK algorithm with the pivot selection strategy that minimizes the number of branches at each node leads to a worst-case optimal algorithm for maximal clique enumeration. 

However, to the best of our knowledge, no other worst-case optimal backtracking algorithms have been developed for enumerating maximal relaxed cliques (Section~\ref{subsec:related-works}). 
This presents two challenges: (1) identifying the worst-case output size, and (2) designing a backtracking algorithm whose search space matches the worst-case output size.

\subject{Novelty of the pivoting}
To address challenge (2), it is essential to develop a new pivoting technique for maximal $k$-defective clique enumeration, as the original pivoting technique can be applied only for maximal clique enumeration. Additionally, a carefully designed criterion for the pivot selection is necessary.
We present a pivoting technique, along with a new pivot selection strategy for our problem. 
This enables our algorithm to explore a worst-case optimal search space when $k$ is a constant.  Challenge (1) will be addressed in Section~\ref{subsec:time-complexity} with the time complexity analysis.

\myparagraph{The pivoting technique}
We first describe the main idea of the pivoting technique.
Consider an instance $I = (S, C, X)$ where $N_C(S) \neq \emptyset$, and let $p$ be a vertex in $N_C(S)$ (i.e., $p$ is adjacent to all vertices in $S$). We refer to $p$ as the \emph{pivot vertex}, and the criterion for selecting it will be explained shortly.
Each partial solution $S'$ generated from $I$, i.e., which includes $S$ and is included by $S \cup C$, falls into one of the following two types: (i) all vertices in $S'$ are adjacent to $p$, or (ii) $S'$ contains $p$ or a non-neighbor of $p$. A partial solution of type (i) is not maximal, as adding $p$ results in another $k$-defective clique. Thus, generating type (i) partial solutions is unnecessary. The pivoting technique ensures that every partial solution generated from $I$ contains either $p$ or a non-neighbor of $p$, thereby avoiding type (i) partial solutions. This is achieved by branching only with $p$ and its non-neighbors.

For ease of presentation, we regard $p$ as a non-neighbor of $p$, i.e., we define $\nnbr_C[p] = \nnbr_C(p) \cup \{p\}$. The following lemma formally introduces our pivoting technique.

\begin{mylemma}[$k$-defective clique pivoting]\label{lemma:pivoting}
Given an instance $(S, C, X)$ of a recursive call where $N_C(S) \neq \emptyset$, let $p$ be a vertex in $N_C(S)$. Every maximal $k$-defective clique that includes $S$ and is included by $S \cup C$ must contain at least one vertex in $\nnbr_C[p]$.
\end{mylemma}

\begin{proof}
    Suppose not. Then, there exists a maximal $k$-defective clique $S'$ that includes $S$ but does not contain any vertex in $\nnbr_C[p]$. Since $p$ is in $N_C(S)$, all the vertices in $S'$ are adjacent to $p$. Therefore, $S' \cup \{p\}$ is also a $k$-defective clique, and thus $S'$ is not maximal. This contradiction proves the lemma.
\end{proof}

Line~\ref{line:pivot} of Algorithm~\ref{alg:backtrack} incorporates the pivoting technique into our branch-and-bound algorithm. If $N_C(S)=\emptyset$, we set $B = C$ since Lemma~\ref{lemma:pivoting} cannot be applied. Otherwise (i.e., $N_C(S) \neq \emptyset$), a pivot vertex $p$ is selected from $N_C(S)$, and we set the set of branching vertices $B$ to $\nnbr_C[p]$. 
In Lines~\ref{line:while}--\ref{line:moving}, recursive calls are made only with the vertices in $B$. Lemma~\ref{lemma:pivoting} guarantees that Algorithm~\ref{alg:backtrack} generates all solutions.

The remaining issue is determining the criterion for selecting a pivot vertex from the vertices in $N_C(S)$ (Line~\ref{line:pivot}). 

\myparagraph{A new pivot selection strategy}
Since we deal with the problem of enumerating $k$-defective cliques, the pivoting strategy of Tomita et al.~\cite{tomita2006worst}, which is designed for maximal clique enumeration, cannot be applied directly. 
We present a new pivot selection strategy for our problem. \redundant{We stress that the proposed pivot selection strategy is instrumental in establishing the worst-case optimality of the algorithm.}{}

Consider an instance $I = (S, C, X)$, where a pivot vertex can be selected (i.e., $N_C(S) \neq \emptyset$). 
Our algorithm selects a pivot vertex $p$ from $N_C(S)$ that has the fewest non-neighbors in $N_C(S)$, i.e.,
$$p = \argmin_{u \in N_C(S)} |\nnbr_{C}[u] \cap N_C(S)|.$$ 

\noindent
Our algorithm has the following characteristics:
\begin{itemize}
    \item Since our branching rule (Line~\ref{line:branching}) first makes branches with the vertices in $B = \nnbr_C[p]$ that are adjacent to all vertices in $S$, it first generates cliques and then adds missing edges.
    \item Our pivoting technique aims to generate maximal solutions as early as possible by branching with $p$'s non-neighbors (and delaying the addition of $p$'s neighbors to $S$).
    \item Since our pivot selection strategy minimizes the number of non-neighbors in $N_C(S)$, it minimizes the number of branches on vertices in $N_C(S)$ (i.e., branches that do not introduce missing edges).
\end{itemize}
These three characteristics together lead to the worst-case optimal search space (Section~\ref{subsec:time-complexity}).
Therefore, this is the first algorithm that is proven to be worst-case optimal among all algorithms for relaxed clique problems.

\begin{myexample}
Figure~\ref{fig:search-tree-ours-pivot} illustrates the search tree generated by Algorithm~\ref{alg:backtrack} with our proposed pivoting technique and pivot selection strategy.
At the root node, $N_C(S) = \{u_1, \ldots, u_8\}$ and the vertex $u_8$ is selected by our pivot selection strategy, since $u_8$ has the fewest non-neighbors in $N_C(S) = \{u_1, \ldots, u_8\}$. As a result, $B$ is set to $\nnbr_C[u_8] = \{u_1, u_8\}$ (Line~\ref{line:pivot}).
Since the first branch with $u_1$ works similarly to Example~\ref{ex:nopivot}, we explain the second branch, $M_1 = (\{u_8\}, \{u_2, u_3, u_4, u_5, u_6, u_7\}, \{u_1\})$. 
Here, $N_C(S) = \{u_2, u_3, u_4, u_5, u_6, u_7\}$, and the numbers of non-neighbors in $N_C(S)$ are 5, 5, 3, 3, 3, and 3 for $u = u_2, u_3, u_4, u_5, u_6$, and $u_7$, respectively. We select $u_7$ as the pivot vertex (we break the tie by selecting the vertex with the highest ID) and
$B$ is set to $\nnbr_C[u_7] = \{u_2, u_3, u_7\}$ in Line~\ref{line:pivot}. It first generates the branch $M_2 = (\{u_8, u_2\}, \{u_3, u_4, u_5, u_6, u_7\}, \{u_1\})$. 
In the subtree rooted at $M_2$, two solutions are output. 
\end{myexample}

\myparagraph{Comparison to existing pivoting technique}
Dai et al.~\cite{dai2023maximal} also proposed a generalized pivoting technique that can be applied to algorithms for enumerating maximal substructures satisfying the hereditary property.
However, this pivoting technique has two limitations: (i) its pivoting technique is applied only when every vertex in $C$ is adjacent to all vertices in $S$ (i.e., $C = N_C(S)$), and (ii) it does not provide criteria for selecting a pivot vertex.

Figure~\ref{fig:search-tree-pivot} shows the search tree of \cite{dai2023maximal} with its pivoting technique. 
The node $M = (\{u_8, u_4\}, \{u_2, u_3, u_5, u_6, u_7\}, \emptyset)$ first branches with $u_3$ and $u_5$ without applying the pivoting technique since they each have a non-neighbor in $\{u_8, u_4\}$. Then, the pivoting technique is applied to $(\{u_8, u_4\}, \{u_2, u_6, u_7\}, \allowbreak \{u_3, u_5\})$. Since there are no criteria for pivot selection, any vertex from $\{u_2, u_6, u_7\}$ can be selected as a pivot, potentially leading to sub-optimal search space.

\subsection{Theoretical Analysis}\label{subsec:time-complexity}
\SetKwFunction{clique}{\bnb{}}

\newcommand{\Ci}[1]{C_{#1}}
\newcommand{\np}[1]{3^{\frac{#1}{3}}}
\newcommand{\ssum}[2]{\sum_{#1}^{#2}}
\newcommand{\pprod}[2]{\prod_{#1}^{#2}}

We now show that the search space of Algorithm~\ref{alg:backtrack} is worst-case optimal when $k$ is a constant.
Since real-world applications typically require small values of $k$ in obtaining $k$-defective cliques~\cite{dai2023maximal}, it is reasonable
to assume that $k$ is a constant.
We first prove the lower bound $\Omega(\np{n} \cdot n^k)$ of the worst-case output size for maximal $k$-defective clique enumeration. Then, we prove the upper bound $\mathcal{O}(\np{n} \cdot n^k$) by analyzing the number of search tree nodes of Algorithm~\ref{alg:backtrack} in the worst case.

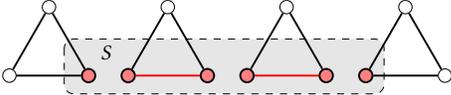
\begin{figure}
    \centering
    \scalebox{0.9}{
        \newcommand{\x}{0.65}
\newcommand{\sq}{1.732}

\begin{tikzpicture}  
  [scale=.9,auto=center,every node/.style={circle, inner sep=0.7mm, draw}, > = stealth] 

  \draw [draw=black, dashed, rounded corners, fill=gray!20] (1*\x - 0.4, -\sq*\x + 0.6) rectangle (8*\x + 0.3, -\sq*\x - 0.3);

  \node [draw=none] (lb) at (1*\x + 0.3, -\sq*\x + 0.38) {$S$};
   
  \node [] (u1) at (0*\x, 0*\x) {};
  \node [] (u2) at (-1*\x, -\sq*\x) {};
  \node [thick, fill=red!50] (u3) at (1*\x, -\sq*\x) {};
  
  \node [] (u4) at (3*\x, 0*\x) {};
  \node [thick, fill=red!50] (u5) at (2*\x, -\sq*\x) {};
  \node [thick, fill=red!50] (u6) at (4*\x, -\sq*\x) {};
  
  \node [] (u7) at (6*\x, 0*\x) {};
  \node [thick, fill=red!50] (u8) at (5*\x, -\sq*\x) {};
  \node [thick, fill=red!50] (u9) at (7*\x, -\sq*\x) {};
  
  \node [] (u10) at (9*\x, 0*\x) {};
  \node [thick, fill=red!50] (u11) at (8*\x, -\sq*\x) {};
  \node [] (u12) at (10*\x, -\sq*\x) {};

  \path[-, thick] (u1) edge (u2); 
  \path[-, thick] (u2) edge (u3); 
  \path[-, thick] (u1) edge (u3); 
  
  \path[-, thick] (u4) edge (u5); 
  \path[-, thick] (u4) edge (u6); 
  \path[-, thick, red] (u5) edge (u6); 
  
  \path[-, thick] (u7) edge (u8); 
  \path[-, thick] (u7) edge (u9); 
  \path[-, thick, red] (u8) edge (u9); 
  
  \path[-, thick] (u10) edge (u11); 
  \path[-, thick] (u10) edge (u12); 
  \path[-, thick] (u11) edge (u12);

\end{tikzpicture}  
    }
    
    \caption{The complement graph of the Moon-Moser graph with 12 vertices. The set $S$, consisting of six vertices and containing two missing edges (highlighted in red), is a maximal 2-defective clique in the Moon-Moser graph.}
    \label{fig:moon-moser}
    \Description{}
\end{figure}

The Moon-Moser graph~\cite{moon1965cliques} is a graph that includes the largest number of maximal cliques among all graphs of the same size. Figure~\ref{fig:moon-moser} shows the complement graph of the Moon-Moser graph with 12 vertices. When the size $n$ of the Moon-Moser graph is a multiple of $3$, the complement graph of the Moon-Moser graph consists of $\frac{n}{3}$ disjoint triangles. 

The following theorem provides the lower bound on the worst-case output size.

\begin{mytheorem}\label{lemma:moonmoser}
    The worst-case output size of maximal $k$-defective clique enumeration is $\Omega(\np{n} \cdot n^k)$ when $k$ is a constant.
\end{mytheorem}

\begin{proof}
    We prove that the Moon-Moser graph with $n$ vertices, where $n$ is a multiple of 3 and $\frac{n}{3} \ge k$, contains at least $3^{\frac{n}{3}} \cdot \binom{n/3}{k}$ maximal $k$-defective cliques; note that $3^{\frac{n}{3}} \cdot \binom{n/3}{k} = \Omega(\np{n} \cdot n^k)$ when $k$ is a constant.
    To construct a maximal $k$-defective clique in the Moon-Moser graph, we first select $k$ triangles in the complement graph of the Moon-Moser graph. Then, we choose a missing edge from each of the $k$ selected triangles and one vertex from each of the remaining $\frac{n}{3} - k$ triangles; for example in Figure~\ref{fig:moon-moser}, we first choose two missing edges from the second and third triangles and then choose two vertices from the remaining triangles. The resulting set of vertices forms a maximal $k$-defective clique in the Moon-Moser graph, since the set contains exactly $k$ missing edges, and adding another vertex to the set introduces a missing edge. The total number of such maximal $k$-defective cliques is $3^{\frac{n}{3}} \cdot \binom{n/3}{k}$.
\end{proof}

Now, we prove an upper bound on the number of nodes in the search tree generated by Algorithm~\ref{alg:backtrack}. Since the outputs (i.e., maximal $k$-defective cliques) appear at leaf nodes in the search tree, the output size is not greater than the number of search tree nodes. Thus, an upper bound on the number of search tree nodes is an upper bound on the output size.

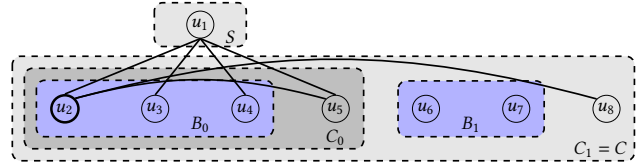
\begin{figure}
    \centering
    \scalebox{0.8}{
        \begin{tikzpicture}[
    node/.style={circle, draw, fill=white, minimum size=4.5mm, inner sep=0pt}, %
    graynode/.style={circle, draw, fill=gray!30, minimum size=4.5mm, inner sep=0pt}, %
    bluenode/.style={circle, draw, fill=blue!30, minimum size=4.5mm, inner sep=0pt}, %
    edge/.style={thick}
]

\newcommand{\sepnode}[0]{1.5}
\newcommand{\hbracket}[0]{0.4}
\newcommand{\hhbracket}[0]{0.6}
\newcommand{\gapratio}[0]{0.5}

\node[node, fill=none, very thick] (n0) at (-1*\sepnode, 0) {$u_2$}; 
\node[node, fill=none] (n1) at (0*\sepnode, 0) {$u_3$}; 
\node[node, fill=none] (n2) at (1*\sepnode, 0) {$u_4$};
\node[node, fill=none] (n3) at (2*\sepnode, 0) {$u_5$};
\node[node, fill=none] (n4) at (3*\sepnode, 0) {$u_6$};
\node[node, fill=none] (n5) at (4*\sepnode, 0) {$u_7$};
\node[node, fill=none] (n6) at (5*\sepnode, 0) {$u_8$};

\node[node, fill=none] (s1) at (0.5*\sepnode, 1.4) {$u_1$}; 
\draw[edge] (s1.south) to (n0.north);
\draw[edge] (s1.south) to (n1.north);
\draw[edge] (s1.south) to (n2.north);
\draw[edge] (s1.south) to (n3.north);

\draw[edge] (n0.north east) to[bend left=15] (n3.north west);
\draw[edge] (n0.north east) to[bend left=15] (n6.north west);

\begin{scope}[on background layer]
    \filldraw[gray!20, rounded corners, dashed, draw=black, thick] ($(n0.north west)+(-0.7,0.7)$) 
        rectangle ($(n6.south east)+(0.3,-0.7)$);
\end{scope}

\begin{scope}[on background layer]
    \filldraw[gray!50, rounded corners, dashed, draw=black, thick] ($(n0.north west)+(-0.5,0.5)$) 
        rectangle ($(n3.south east)+(0.3,-0.5)$);
\end{scope}

\begin{scope}[on background layer]
    \filldraw[blue!30, rounded corners, dashed, draw=black, thick] ($(n0.north west)+(-0.3,0.3)$) 
        rectangle ($(n2.south east)+(0.3,-0.3)$);
\end{scope}

\begin{scope}[on background layer]
    \filldraw[blue!30, rounded corners, dashed, draw=black, thick] ($(n4.north west)+(-0.3,0.3)$) 
        rectangle ($(n5.south east)+(0.3,-0.3)$);
\end{scope}

\begin{scope}[on background layer]
    \filldraw[gray!20, rounded corners, dashed, draw=black, thick] ($(s1.north west)+(-0.6,0.2)$) 
        rectangle ($(s1.south east)+(0.6,-0.2)$);
\end{scope}

\node[anchor=north] at ($(n1.south east)!0.5!(n2.south west) + (0,0.15)$) {$B_0$};
\node[anchor=north] at ($(n4.south east)!0.5!(n5.south west) + (0,0.15)$) {$B_1$};
\node[anchor=north] at ($(n3.south) + (0,0.0)$) {$C_0$};
\node[anchor=north] at ($(n6.south) + (-0.1,-0.2)$) {$C_1=C$};
\node[anchor=north] at ($(s1.south) + (0.5,0.25)$) {$S$};

\end{tikzpicture}

\newcommand{\nlabel}[4]{#4}
\newcommand{\nlabelp}[5]{#5}
\newcommand{\nodefec}[1]{$u_{#1}$}
\newcommand{\defec}[1]{$\textcolor{red}{u_{#1}}$}
    }
    
    \caption{A diagram illustrating the definitions of $C_i$ and $B_i$ for $I = (S, C, X)$, where $S = \{u_1\}$, $C = \{u_2, \ldots, u_8\}$, and $\kappa = k = 1$. The vertex $u_2$ is selected as the pivot vertex, and $B$ is set to $\nnbr_C[u_1] = \{u_2, u_3, u_4, u_6, u_7\}$.}
    \label{fig:proof-illust}
    \Description{}
\end{figure}

To formally establish the upper bound, we introduce some notations. 
Let $\mathcal{T}$ be the search tree generated by Algorithm~\ref{alg:backtrack}. 
Suppose that we are given an instance $(S, C, X)$ of $\mathcal{T}$. We denote $T(S, C, X)$ by the number of nodes in the subtree of $\mathcal{T}$ rooted at $(S, C, X)$. Let $\kappa$ be the number of missing edges that can be added to $S$ (i.e., $\kappa = k - |\overline{E}(S)|$). 
For every integer $0 \le i \le \kappa$, let $C_i$ be the set of vertices in $C$ that introduce at most $i$ missing edges when added to $S$ (i.e., $C_i = \{u \in C: |\nnbr_S(u)| \le i\}$); see Figure~\ref{fig:proof-illust} for illustration (for now, ignore $B_i$). For $\kappa$ non-negative integers $n_1, \ldots, n_{\kappa}$, we define 
$$f^{\kappa}(n_1, \ldots, n_{\kappa}) = 1 + n_{\kappa} + n_{\kappa - 1}n_{\kappa} + \cdots +  n_1 \cdots n_{\kappa} = 1 + \ssum{i=1}{\kappa} \pprod{j=i}{\kappa} n_j.$$

We prove an upper bound of $T(S, C, X)$ in the following lemma.

\newenvironment{proofsketch}{%
  \renewcommand{\proofname}{Proof Sketch}\proof}{\endproof}

\begin{mylemma}\label{lemma:inequality}
    For any instance $I = (S, C, X)$ of the search tree $\mathcal{T}$,
    \begin{align}
        T(S, C, X) &\le 2 \cdot \np{|C_0|} \cdot f^{\kappa}(|C_1|, \ldots, |C_{\kappa}|). \label{eq:ineq}
    \end{align}
\end{mylemma}

\begin{proof}
    We prove the inequality by induction. For the base case where $I$ is a leaf node, it is trivial that $T(S, C, X) = 1 \le 2 \cdot 3^{\frac{|C_0|}{3}} \cdot f^{\kappa}(|C_1|, \ldots, |C_{\kappa}|) = 2$ because $C_0 \subseteq C_1 \subseteq \cdots \subseteq C_{\kappa} = C = \emptyset$ (that is, $|C_0| = |C_1| = \cdots = |C_{\kappa}| = 0$). We now consider an arbitrary non-leaf node $I$. There are two cases as follows.  

Case I: $C_0 = \emptyset$. In this case, we can observe that adding a candidate vertex in $C$ to the partial solution always increases the number of missing edges. 
From the observation, the height of the subtree of $\mathcal{T}$ rooted at $I$ is bounded by $\kappa$.
Let the depth of the root node of the subtree be zero. 
Consider a node $I$ in the subtree at depth $i$ ($0 \le i < \kappa$). Then, the number of allowed missing edges at $I$ is at most $\kappa - i$ from the observation, and branches at $I$ can be made only with vertices in $C_{\kappa - i}$. Therefore, the number of nodes at depth $i + 1$ is at most $|C_{\kappa - i}|\cdots |C_{\kappa}|$, and thus the total number of nodes in the subtree is at most $1 + |C_{\kappa}| + |C_{\kappa - 1}||C_{\kappa}| + \cdots + |C_1| \cdots |C_{\kappa}| = f^{\kappa}(|C_1|, \ldots, |C_{\kappa}|)$. 

Case II: $C_0 \neq \emptyset$. When $\kappa = 0$ (i.e., no missing edge is allowed), the problem degenerates to maximal clique enumeration. In this case, our pivot selection strategy coincides with that of Tomita et al.~\cite{tomita2006worst}, and therefore we obtain $T(S, C, X) \le 2 \cdot \np{|C_0|}$~\cite{fomin2010exact}. Thus, we consider the case $\kappa > 0$.

Since $C_0 = N_C(S)$, our pivot selection strategy selects a pivot vertex $p \in N_C(S)$ that has the fewest non-neighbors in $N_C(S)$, and then sets $B$ to $\nnbr_C[p]$ (Line~\ref{line:pivot}). For every integer $0 \le i \le \kappa$, let $B_i$ be the set of vertices in $B$ that introduce exactly $i$ missing edges when added to $S$ (i.e., $B_i = \{u \in B: |\nnbr_S(u)| = i\}$); see Figure~\ref{fig:proof-illust} for illustration.

We refer to an instance of a recursive call from $I$ as a \emph{child instance} of $I$.
Consider $I' = (S' = S \cup \{b\}, C', X')$, a child instance of $I$ with the branching vertex $b \in B = B_0 \cup B_1 \cup \cdots \cup B_{\kappa}$. Define $\kappa' = \kappa - |\nnbr_S(b)|$, the number of missing edges that can be added to $S' = S \cup \{b\}$. For $0 \le j \le \kappa'$, define $C'_j = \{u \in C': |\nnbr_{S'}(u)| \le j\}$, the set of vertices in $C'$ that introduce at most $j$ missing edges when added to $S'$. For example, in Figure~\ref{fig:proof-illust}, consider a child instance of $I$ with the branching vertex $u_2$, which results in $(S', C', X') = (\{u_1, u_2\}, \{u_3, u_4, u_5, u_8\}, \emptyset)$. In this instance, we have $\kappa' = 1$, $C'_0 = \{u_5\}$, and $C'_1 = \{u_3, u_4, u_5, u_8\}$.

We assume as the induction hypothesis that the following inequality holds for every child instance $I' = (S', C', X')$ of $I$:$$T(S', C', X') \le 2 \cdot 3^{\frac{|C'_0|}{3}} \cdot f^{\kappa'}(|C'_1|, \ldots, |C'_{\kappa'}|).$$ We now show that inequality~(\ref{eq:ineq}) holds for $I$ where $C_0 \neq \emptyset$ and $\kappa > 0$. Every child instance $I' = (S' = S \cup \{b\}, C', X')$ of $I$ falls into one of the following three types.
\begin{enumerate}[label=(\alph*), leftmargin=5mm]
    \item When $b = p$:
    \begin{itemize}
        \item we have $\kappa' = \kappa - |\nnbr_S(b)| = \kappa$ because $p \in B_0$.
        \item Since $b = p$ is in $S'$, we have $p \notin C'_j$ for every $0 \le j \le \kappa$. Every vertex in $B_0 \setminus \{p\}$ does not belong to $C'_0$ because it has one non-neighbor in $S' = S \cup \{p\}$. Thus, we obtain $C'_0 \subseteq C_0 \setminus B_0$.
        \item Similarly, for $1 \le j \le \kappa$, every vertex in $B_j$ does not belong to $C'_j$ because it has $j+1$ non-neighbors in $S' = S \cup \{p\}$. Thus, we obtain $C'_j \subseteq C_j \setminus B_j \setminus \{p\}$ for $1 \le j \le \kappa$.
        \item Putting everything together, we obtain $T(S', C', X') \le 2 \cdot 3^{\frac{|C_0 \setminus B_0|}{3}} \cdot f^{\kappa}(|C_1 \setminus B_1 \setminus \{p\}|, \ldots, |C_{\kappa} \setminus B_{\kappa} \setminus \{p\}|)$.
    \end{itemize}
    \label{item:aa}
    
    \item When $b \in B_0 \setminus \{p\}$:
    \begin{itemize}
        \item we have $\kappa' = \kappa$.
        \item We now show that $|C'_0| \le |C_0 \setminus B_0|$ holds. 
        We have $C'_0 \subseteq C_0 \setminus \nnbr_{C_0}[b]$ because $b \in S'$ and every non-neighbor in $\nnbr_C(b)$ has one non-neighbor in $S'$.
        Since $C_0 = N_C(S)$, by our pivot selection strategy, we have $|\nnbr_{C_0}[b]| = |\nnbr_C[b] \cap N_C(S)| \ge |\nnbr_C[p] \cap N_C(S)| = |B_0|$.
        Thus, we obtain $|C'_0| \le |C_0 \setminus \nnbr_{C_0}[b]| = |C_0| - |\nnbr_{C_0}[b]| \le |C_0| - |B_0| = |C_0 \setminus B_0|$.
        \item As a result, $T(S', C', X') \le 2 \cdot 3^{\frac{|C_0 \setminus B_0|}{3}} \cdot f^{\kappa}(|C_1|, \ldots, |C_{\kappa}|)$.
    \end{itemize}
    \label{item:bb}
    
    \item When $b \in B_i$ $(1 \le i \le \kappa)$:
    \begin{itemize}
        \item we have $\kappa' = \kappa - i$ because $b \in B_i$.
        \item According to our branching rule in Line~\ref{line:branching} of Algorithm~\ref{alg:backtrack}, when branching with $b \notin B_0 = N_B(S)$, all child instances with branching vertices in $B_0$ have already been processed and removed from $C$ in Line~\ref{line:moving}. That is, every vertex in $B_0$ is excluded from $C'$. Thus, for every $0 \le j \le \kappa - i$, we obtain $C'_j \subseteq C_j \setminus B_0$.
        \item Thus, $T(S', C', X') \le 2 \cdot 3^{\frac{|C_0 \setminus B_0|}{3}} \cdot f^{\kappa - i}(|C_1 \setminus B_0|, \ldots, |C_{\kappa - i} \setminus B_0|)$.
    \end{itemize}
    \label{item:cc}
\end{enumerate}

The numbers of type~\ref{item:aa}, type~\ref{item:bb}, and type~\ref{item:cc} branches are $1$, $|B_0| - 1$, and $|B_i|$, respectively. For brevity, we denote $|C_i| = n_i$ and $|B_i| = y_i$ for $0 \le i \le \kappa$.
Then, we obtain the following inequality:
\begin{align*}
    T(S, C, X)
    =&~ \sum_{\text{every child instance $(S', C', X')$ of $I$}} T(S', C', X') + 1  \\
    \le&~ 2 \cdot 3^{\frac{n_0 - y_0}{3}} \cdot f^{\kappa}(n_1 - y_1 - 1, \ldots, n_{\kappa} - y_{\kappa} - 1)   \\
    +&~ (y_0 - 1) \cdot 2 \cdot 3^{\frac{n_0 - y_0}{3}} \cdot f^{\kappa}(n_1, \ldots, n_{\kappa})  \\
    +&~ \sum_{i=1}^{\kappa} y_i \cdot 2 \cdot 3^{\frac{n_0 - y_0}{3}} \cdot f^{\kappa-i}(n_1 - y_0, \ldots, n_{\kappa-i} - y_0)  \\
    +&~ 1.
\end{align*}

We now show that the right-hand side of the above inequality is at most $2 \cdot \np{|C_0|} \cdot f^{\kappa}(|C_1|, \ldots, |C_{\kappa}|) = 2 \cdot 3^{\frac{n_0}{3}} \cdot f^{\kappa}(n_1, \ldots, n_{\kappa})$. We modify the second term on the right-hand side as follows:
\begin{align*}
    &~ (y_0 - 1) \cdot 2 \cdot 3^{\frac{n_0 - y_0}{3}} \cdot f^{\kappa}(n_1, \ldots, n_{\kappa}) \\
    =&~ 2 \cdot 3^{\frac{n_0}{3}} \cdot 3^{-\frac{y_0}{3}} \cdot y_0 \cdot f^{\kappa}(n_1, \ldots, n_{\kappa}) - 2 \cdot 3^{\frac{n_0 - y_0}{3}} \cdot f^{\kappa}(n_1, \ldots, n_{\kappa})  \\
    \le&~ 2 \cdot 3^{\frac{n_0}{3}} \cdot f^{\kappa}(n_1, \ldots, n_{\kappa}) - 2 \cdot 3^{\frac{n_0 - y_0}{3}} \cdot f^{\kappa}(n_1, \ldots, n_{\kappa}).
\end{align*}
The last inequality holds because $3^{-\frac{y_0}{3}} \cdot y_0 \le 1$ for any positive integer $y_0$~\cite{tomita2006worst}.

Thus, the following inequality holds:
\begin{align*}
   T(S, C, X) 
    \le&~ 2 \cdot 3^{\frac{n_0}{3}} \cdot f^{\kappa}(n_1, \ldots, n_{\kappa}) + 2 \cdot 3^{\frac{n_0 - y_0}{3}} \cdot (\alpha + \beta - \gamma) + 1
\end{align*}
where
\begin{align*}
    \alpha &= f^{\kappa}(n_1 - y_1 - 1, \ldots, n_{\kappa} - y_{\kappa} - 1) = 1 + \ssum{i=1}{\kappa}\pprod{j=i}{\kappa}(n_j - y_j - 1), \\ 
    \beta &= \ssum{i=1}{\kappa} y_i \cdot f^{\kappa-i}(n_1 - y_0, \ldots, n_{\kappa-i} - y_0)\text{, and} \\
    \gamma &= f^{\kappa}(n_1, \ldots, n_{\kappa}) = 1 + \ssum{i=1}{\kappa}\pprod{j=i}{\kappa}n_j
\end{align*}
($\alpha$ and $\beta$ are derived from the first and third terms, respectively).

Now, it suffices to show that $2 \cdot 3^{\frac{n_0 - y_0}{3}} \cdot (\alpha + \beta - \gamma) + 1 \le 0$. 
To obtain this, a little manipulation is needed.
Let $\alpha_i = \pprod{j=i}{\kappa}(n_j - y_j - 1)$, $\beta_i = y_i \cdot f^{\kappa - i}(n_1 - y_0, \ldots, n_{\kappa - i} - y_0)$, and $\gamma_i = \pprod{j=i}{\kappa} n_j$. Then, we can rewrite
\begin{align*}
    2 \cdot 3^{\frac{n_0 - y_0}{3}} \cdot (\alpha + \beta - \gamma) + 1 
    &= 2 \cdot 3^{\frac{n_0 - y_0}{3}} \cdot \ssum{i=1}{\kappa}(\alpha_i + \beta_i - \gamma_i) + 1 \\
    &= 2 \cdot 3^{\frac{n_0 - y_0}{3}} \cdot \ssum{i=1}{\kappa - 1}(\alpha_i + \beta_i - \gamma_i)  \\ 
    &+ 2 \cdot 3^{\frac{n_0 - y_0}{3}} \cdot (\alpha_{\kappa} + \beta_{\kappa} - \gamma_{\kappa}) + 1 \\ 
    &< 2 \cdot 3^{\frac{n_0 - y_0}{3}} \cdot \ssum{i=1}{\kappa - 1}(\alpha_i + \beta_i - \gamma_i).
\end{align*}
The last inequality holds because $\alpha_{\kappa} + \beta_{\kappa} - \gamma_{\kappa} = (n_{\kappa} - y_{\kappa} - 1) + y_{\kappa} - n_{\kappa} =  -1$ and $n_0 - y_0 = |C_0 \setminus B_0| \ge 0$.

Finally, we conclude the lemma by showing that $\alpha_i + \beta_i - \gamma_i \le 0$ for $1 \le i\le \kappa - 1$. We first obtain the following inequality by a simple manipulation:
\begin{align*}
    \alpha_i - \gamma_i &= \pprod{j=i}{\kappa}(n_j - y_j - 1) - \pprod{j=i}{\kappa} n_j \\
    &\le \pprod{j=i}{\kappa}(n_j - y_j) - \pprod{j=i}{\kappa} n_j \\
    &= \pprod{j=i}{\kappa}(n_j - y_j) - \pprod{j=i}{\kappa} n_j + y_i \cdot \pprod{j=i+1}{\kappa}n_j - y_i \cdot \pprod{j=i+1}{\kappa}n_j\\
    &= (n_i - y_i) \cdot \left[ \pprod{j=i+1}{\kappa}(n_j - y_j) - \pprod{j=i+1}{\kappa}n_j \right] - y_i \cdot \pprod{j=i+1}{\kappa}n_j \\
    &\le -y_i \cdot \pprod{j=i+1}{\kappa}n_j,
\end{align*}
where $n_j - y_j \ge 1$ because $n_j - y_j = |C_j \setminus B_j|$ and $p \in C_j \setminus B_j$.
Moreover, we have $\beta_i = y_i \cdot f^{\kappa - i}(n_1 - y_0, \ldots, n_{\kappa - i} - y_0) \le y_i \cdot f^{\kappa - i}(n_{i+1} - 1, \ldots, n_{\kappa} - 1)$ because $n_1 \le n_2 \le \cdots \le n_{\kappa}$ and $y_0 = |B_0| \ge 1$.
Thus, we obtain the following inequality:
\begin{align*}
    \alpha_i + \beta_i - \gamma_i 
    &\le y_i \cdot f^{\kappa - i}(n_{i+1} - 1, \ldots, n_{\kappa} - 1) - y_i \cdot \pprod{j=i+1}{\kappa}n_j \\
    &= y_i \cdot \left[1 + \ssum{s=1}{\kappa-i} \pprod{j=i+s}{\kappa} (n_{j} - 1) - \pprod{j=i+1}{\kappa}n_j \right] \\
    &\le y_i \cdot \left[1 + \ssum{s=1}{\kappa-i} 
    \left\{ (n_{i+s} - 1) \cdot \pprod{j=i+s+1}{\kappa} n_{j} \right\} - \pprod{j=i+1}{\kappa}n_j \right] 
    \tag{Because $\pprod{j=i+s+1}{\kappa} (n_{j} - 1) \le \pprod{j=i+s+1}{\kappa} n_{j}$} \\
    &= y_i \cdot \left[1 + \ssum{s=1}{\kappa-i} \pprod{j=i+s}{\kappa} n_{j} - \ssum{s=1}{\kappa-i} \pprod{j=i+s+1}{\kappa} n_{j} - \pprod{j=i+1}{\kappa} n_j \right] \\
    &= y_i \cdot \left[1 + \ssum{s=1}{\kappa-i} \pprod{j=i+s}{\kappa} n_{j} - \ssum{s=2}{\kappa-i+1} \pprod{j=i+s}{\kappa} n_{j} - \pprod{j=i+1}{\kappa} n_j \right] \\
    &= 0. \tag{By telescoping}
\end{align*}

Therefore, the lemma is proved.
\end{proof}

From Lemma~\ref{lemma:inequality}, we obtain an upper bound on the number of search tree nodes generated by Algorithm~\ref{alg:backtrack}.
\begin{mytheorem}\label{thm:time1}
    Given a graph $G=(V, E)$ with $n$ vertices and a non-negative integer $k$, the number of search tree nodes generated by Algorithm~\ref{alg:backtrack} is $\mathcal{O}(3^{\frac{n}{3}} \cdot n^k)$.    
\end{mytheorem}

\begin{proof}
    Consider the root node $(S, C, X) = (\emptyset, V, \emptyset)$ of the search tree. Since $S = \emptyset$, we have $C_0 = C_1 = \cdots = C_k = V$.
    Therefore, $T(\emptyset, V, \emptyset) \le 2 \cdot 3^{\frac{n}{3}} \cdot f^k(n, \ldots, n) = 2 \cdot 3^{\frac{n}{3}} \cdot (1 + n + n^{2} + \cdots + n^{\kappa}) = \mathcal{O}(\np{n} \cdot n^{k})$.
\end{proof}

Combined with Theorem~\ref{lemma:moonmoser}, the search space size of Algorithm~\ref{alg:backtrack} matches the worst-case output size $\Omega(3^{\frac{n}{3}} \cdot n^k)$ of the problem when $k$ is a constant. 

From Theorem~\ref{thm:time1}, the following theorem holds since each search tree node takes $\mathcal{O}(m)$ time.
\begin{mytheorem}\label{thm:time-complexity1}
    Algorithm~\ref{alg:backtrack} runs in $\mathcal{O}(m \cdot 3^{\frac{n}{3}} \cdot n^k)$ time.
\end{mytheorem}

\begin{proof}
    In Algorithm~\ref{alg:backtrack}, Line~\ref{line:pivot} can be done in $\mathcal{O}(m)$ time.
    For each branching vertex, the upper bound in Line~\ref{line:ub2} can be computed in $\mathcal{O}(m)$ time~\cite{chang2023efficient}. Additionally, Lines~\ref{line:branching}, \ref{line:refine}, and \ref{line:moving} also run in $\mathcal{O}(m)$ time. That is, each search tree node takes $\mathcal{O}(m)$ time. Since Algorithm~\ref{alg:backtrack} generates at most $\mathcal{O}(3^{\frac{n}{3}} \cdot n^k)$ search tree nodes, its total running time is $\mathcal{O}(m \cdot 3^{\frac{n}{3}} \cdot n^k)$.
\end{proof}

\subsection{Using Diameter-Two Property}\label{subsec:twohop}
\subject{When $q \ge k+2$, two-hop neighbors reduction.
Time complexity: $\mathcal{O}^*(n \cdot (\alpha \delta)^{k-1} \cdot 3^{\frac{\delta}{3}})$
}

In this subsection, we describe how we can further reduce the exponent of the time complexity of our algorithm based on the diameter-two property (Property~\ref{prop:diam2}) when $q \ge k + 2$.

The main idea is to decompose the original problem into small subproblems; for each $v_i \in V$, enumerate every solution whose smallest vertex (with respect to a degeneracy ordering) is $v_i$ and whose diameter is at most two, as in \cite{conte2018d2k}. Algorithm~\ref{alg:diameter-two} presents this approach. We first compute a degeneracy ordering $(v_1, v_2, \ldots, v_n)$ of $G$ in Line~\ref{line3:degen}. 
For each vertex $v_i \in V$, we compute the set $V_i = N(v_i) \cup \bigcup_{v \in N^+(v_i)} N(v)$ (Line~\ref{line3:V}). This set contains all solutions of diameter at most two, each of which contains $v_i$ and a forward neighbor of $v_i$~\cite{dai2022scaling}. By the diameter-two property, all vertices in any solution of the subproblem, except for $v_i$, must exist in $V_i \cap \{v_{i + 1}, v_{i + 2}, \ldots, v_n\}$~\cite{conte2018d2k}. Therefore, we set $C_i = V_i \cap \{v_{i + 1}, v_{i + 2}, \ldots, v_n\}$ (Line~\ref{line3:C}). Then, we set $X_i = V_i \cap \{v_1, v_2, \ldots, v_{i - 1}\}$ (Line~\ref{line3:X}).
Finally, a recursive call $(\{v_i\}, C_i, X_i)$ is made (Line~\ref{line3:branch}) to enumerate all solutions of the subproblem.

\begin{algorithm}[t]
    \caption{Our branch-and-bound algorithm utilizing the diameter-two property}
    \label{alg:diameter-two}
    \SetKwProg{myproc}{Procedure}{}{}
\SetKwFunction{branchAndBound}{backtrack}
\SetKwFunction{clique}{\bnb{}}
\SetKwFunction{expand}{ExtendDefective}
\SetKwInOut{KwIn}{Input}
\SetKwInOut{KwOut}{Output}
\SetKw{Continue}{continue}
\SetKw{output}{output}
\SetKw{report}{report}
\KwIn{A graph $G=(V, E)$ and two integers $k$, $q \ge k+2$}
\KwOut{Every maximal $k$-defective clique in $G$ of size $\ge q$}
Let $(v_1, v_2, \ldots, v_n)$ be a degeneracy ordering of $G$\;\label{line3:degen}
\ForEach{$v_i \in V$\label{line3:foreach}}{
    $V_i \gets N(v_i) \cup \bigcup_{v \in N^+(v_i)}{N(v)}$\;\label{line3:V}
    $C_i \gets V_i \cap \{v_{i+1}, v_{i+2}, \ldots, v_n\}$\;\label{line3:C}
    $X_i \gets V_i \cap \{v_1, v_2, \ldots, v_{i-1}\}$\;\label{line3:X}
    \clique{$\{v_i\}, C_i, X_i$}\;\label{line3:branch}
}

\end{algorithm}

\begin{mylemma}\label{lemma:inequality2}
    The number of search tree nodes generated by Algorithm~\ref{alg:diameter-two} is bounded by $\mathcal{O}(n \cdot 3^{\frac{\delta}{3}} \cdot (\delta \Delta)^{k})$.
\end{mylemma}

\begin{proof}
    We prove the lemma by showing that the search space size of each subproblem is bounded by $\mathcal{O}(3^{\frac{\delta}{3}} \cdot (\delta \Delta)^k)$. This follows from the facts that $|N_{C_i}(v_i)| \le \delta$ and $|C_i| \le \delta\Delta$~\cite{chang2024maximum} for all $1 \le i \le n$. Then, applying Lemma~\ref{lemma:inequality}, the overall bound follows.
\end{proof}

Since each search tree node takes $\mathcal{O}((\delta \Delta)^2)$ time, the following theorem holds.

\begin{mytheorem}\label{thm:time-complexity2}
    Algorithm~\ref{alg:diameter-two} runs in $\mathcal{O}(n \cdot 3^{\frac{\delta}{3}} \cdot (\delta \Delta)^{k+2})$ time.
\end{mytheorem}

\begin{proof}
    In Algorithm~\ref{alg:diameter-two}, the running time of Line~\ref{line3:degen} is bounded by $\mathcal{O}(m)$. Additionally, for each branching vertex $v_i$, the total running time of Lines~\ref{line3:V}--\ref{line3:X} is bounded by the number of edges in $G[V_i]$. Since we have $|V_i| \le |N(v_i)| + |N^+(v_i)| \cdot \Delta \le \Delta + \delta \Delta = \mathcal{O}(\delta \Delta)$, the number of edges in $G[V_i]$ is bounded by $\mathcal{O}((\delta \Delta)^2)$. Furthermore, the recursive call in Line~\ref{line3:branch} runs in $\mathcal{O}(3^{\frac{\delta}{3}} \cdot (\delta \Delta)^{k + 2})$ time because the number of search tree nodes rooted at $(\{v_i\}, C_i, X_i)$ is $\mathcal{O}(3^{\frac{\delta}{3}} \cdot (\delta \Delta)^k)$, and each search tree node takes $\mathcal{O}((\delta \Delta)^2)$ time. Therefore, the total running time is $\mathcal{O}(n \cdot 3^{\frac{\delta}{3}} \cdot (\delta \Delta)^{k + 2})$.
\end{proof}

\section{Maximum \texorpdfstring{$k$}{k}-Defective Clique Search}\label{sec:preprocessing}

In this section, we introduce our framework for solving maximum $k$-defective clique search by using our branch-and-bound algorithm for maximal $k$-defective clique enumeration. This framework also incorporates new practical techniques to reduce the search space: (1) a technique for computing a large initial solution, and (2) a graph reduction technique for eliminating unpromising vertices and edges from the input graph.

\subsection{Our Framework}

Algorithm~\ref{alg:maximum} presents our framework. 
Given an input graph $G$ and an integer $k$, it first computes an \emph{initial solution} (Section~\ref{subsec:bounding}), which is a $k$-defective clique, and sets $S^*$ to this initial solution. Throughout the execution, $S^*$ maintains the largest $k$-defective clique found so far.  Then, $q$ is set to $|S^*| + 1$ (Line~\ref{line4:setq}); since we have already found $S^*$, we aim to find a $k$-defective clique of size at least $q = |S^*| + 1$. The algorithm then computes a \emph{reduced graph} $g$ (Section~\ref{subsec:reduction}), by removing vertices and edges from $G$ that cannot be contained in any $k$-defective clique of size at least $q$ (Line~\ref{line4:reduce}). If $g$ is empty (i.e., it contains no vertices and edges), then $G$ does not contain any $k$-defective clique of size at least $q$. Therefore, the algorithm returns $S^*$ as a maximum $k$-defective clique (Line~\ref{line4:return}).
If $g$ is not empty, the framework finds a maximum $k$-defective clique using our branch-and-bound algorithm (Lines~\ref{line4:alg2}--\ref{line4:alg1}).
During the branch-and-bound search, whenever a $k$-defective clique $S$ of size at least $q$ (that is, larger than $S^*$) is found, we update $S^*$ to $S$ and set $q$ to $|S^*| + 1$ (Lines~\ref{line4:sizecheck}--\ref{line4:setq2}).
This approach is highly efficient because a larger $q$ allows us to prune more of the search space.

\begin{algorithm}[t]
    \caption{Our framework for maximum $k$-defective clique search}
    \label{alg:maximum}
    \SetKwProg{myproc}{Procedure}{}{}
\SetKwFunction{clique}{\bnb{}}
\SetKwFunction{cliquemaximum}{\bnb{}-maximum}
\SetKwFunction{reduce}{reduce}
\SetKwFunction{heuristic}{initial-solution}
\SetKwInOut{KwIn}{Input}
\SetKwInOut{KwOut}{Output}
\SetKw{Continue}{continue}
\SetKw{output}{output}
\SetKw{report}{report}
\SetKwFunction{update}{refine}
\SetKwFunction{upperbound}{upper-bound}
\KwIn{A graph $G=(V, E)$ and an integer $k$}
\KwOut{A maximum $k$-defective clique in $G$}
$S^* \gets$ \heuristic{$G, k$};
$q \gets |S^*| + 1$\;\label{line4:setq}
$g \gets$ \reduce($G, k, q$)\;\label{line4:reduce}
\lIf{$g$ is empty}{\Return{$S^*$}}\label{line4:return}
\lIf{$q \ge k + 2$}{run Algorithm~\ref{alg:diameter-two} with $g$, $k$, and $q$, in which \clique is replaced with \cliquemaximum}\label{line4:alg2}
\lElse{run Algorithm~\ref{alg:backtrack} with $g$, $k$, and $q$, in which \clique is replaced with \cliquemaximum}\label{line4:alg1}
\Return{$S^*$}\;\label{line4:return2}
\myproc{\cliquemaximum{$S, C, X$}}{
    \If{$|S| \ge q$}{\label{line4:sizecheck}
        $S^* \gets S$; 
        $q \gets |S^*| + 1$\; \label{line4:setq2}
    }
    Same as Lines~\ref{line:pivot}--\ref{line:moving} in Algorithm~\ref{alg:backtrack}, except that \clique in Line~\ref{line:recursive} is replaced with \cliquemaximum\;
}

\end{algorithm}

\subsection{Initial Solution and Graph Reduction}\label{subsec:practical-techniques}
We now present our practical techniques for initial solution and graph reduction.
To introduce our techniques, we first present the concepts of colorful $s$-core and colorful degeneracy ordering, which are introduced in \cite{pan2022fairness, zhang2023fairness}.

\begin{figure}
    \centering
    \scalebox{0.8}{
        \newcommand{\x}{1.0}

\begin{minipage}{.35\textwidth}
\centering
\begin{tikzpicture}  
  [scale=.9,auto=center,every node/.style={circle, inner sep=0.5mm, draw}, > = stealth] 
  \node [fill=yellow!40] (u1) at (4.2*\x, 0.6*\x) {$u_1$};
  \node [fill=blue!40] (u2) at (4.2*\x, -0.6*\x) {$u_2$};
  \node [fill=red!40] (u3) at (3*\x, 1*\x) {$u_3$};
  \node [fill=red!40] (u4) at (3*\x, -1*\x) {$u_4$};
  \node [fill=green!40] (u5) at (2*\x, 0*\x) {$u_5$};
  \node [fill=blue!40] (u6) at (1*\x, 1*\x) {$u_6$};
  \node [fill=blue!40] (u7) at (1*\x, -1*\x) {$u_7$};
  \node [fill=green!40] (u8) at (0*\x, 0*\x) {$u_8$};
  \node [fill=red!40] (u9) at (1*\x, 0*\x) {$u_9$};

  \draw[-, thick] (u3) to[bend right=8] (u8);
  \draw[-, thick] (u4) to[bend left=8] (u8);
  \path[-, thick] (u8) edge (u6); 
  \path[-, thick] (u8) edge (u7); 
  \path[-, thick] (u6) edge (u3); 
  \path[-, thick] (u6) edge (u5); 
  \path[-, thick] (u7) edge (u4); 
  \path[-, thick] (u7) edge (u5); 
  \path[-, thick] (u3) edge (u1); 
  \path[-, thick] (u3) edge (u2); 
  \path[-, thick] (u3) edge (u5); 
  \path[-, thick] (u4) edge (u1); 
  \path[-, thick] (u4) edge (u2); 
  \path[-, thick] (u4) edge (u5); 
  \path[-, thick] (u1) edge (u2); 
  \path[-, thick] (u1) edge (u5); 
  \path[-, thick] (u2) edge (u5); 
  \path[-, thick] (u9) edge (u8); 
  \path[-, thick] (u9) edge (u6); 
  \path[-, thick] (u9) edge (u7); 
  \path[-, thick] (u9) edge (u5); 
\end{tikzpicture}  
\end{minipage}

\begin{minipage}{.2\textwidth}
\centering
\begin{tabular}{c|c}
    \toprule
    Vertices & $\Color(\cdot)$ \\ \midrule
    $u_3, u_4, u_9$ & 1 \\ \hline
    $u_5, u_8$ & 2 \\ \hline
    $u_2, u_6, u_7$ & 3 \\ \hline
    $u_1$ & 4 \\ \bottomrule
\end{tabular}

\end{minipage}
    }

    \caption{An example graph and its coloring $\Color$ (best viewed in color) for applying our practical techniques.}
    \label{fig:input-graph-preprocessing}
\end{figure}
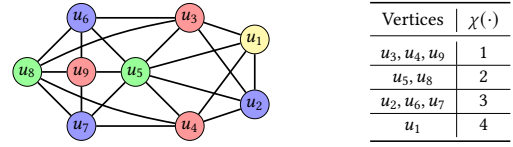

We first color the input graph. Figure~\ref{fig:input-graph-preprocessing} shows a new running example with graph coloring $\chi$.

\begin{mydefinition}[Colorful degree]
Given a graph $G=(V, E)$ and a coloring $\Color$ of $G$, the \emph{colorful degree} of $u\in V$ in $G$ with respect to $\Color$, denoted by $d^{\Color}_G(u)$, is the number of distinct colors among $u$'s neighbors. Formally, $d^{\Color}_G(u) = |\{ \Color(v): v \in N_G(u) \}|$.
\end{mydefinition}

In the graph of Figure~\ref{fig:input-graph-preprocessing}, vertices $u_1, u_2, \ldots, u_5$ have a colorful degree of 3, and the remaining vertices have a colorful degree of $2$ with respect to $\chi$.

\begin{mydefinition}[Colorful $s$-core]
Given a graph $G=(V, E)$, a coloring $\Color$ of $G$ and an integer $s$, the \emph{colorful $s$-core} of $G$ with respect to $\Color$ is the maximal subgraph $g$ of $G$ in which every vertex $u$ in $g$ has colorful degree $d^{\Color}_{g}(u) \ge s$.
\end{mydefinition}

The graph itself in Figure~\ref{fig:input-graph-preprocessing} is the colorful 2-core, and the subgraph induced by $\{u_1, u_2, u_3, u_4, u_5\}$ is the colorful 3-core of the graph. The colorful 4-core of the graph is an empty graph. 

\begin{mydefinition}[Colorful degeneracy ordering]
Given a graph $G=(V, E)$ and a coloring $\Color$ of $G$, an ordering $(v_1, v_2, \ldots, v_n)$ of its vertices $V$ is a \emph{colorful degeneracy ordering} with respect to $\Color$ if, for each $1 \le i \le n$, $v_i$ is a vertex with the smallest colorful degree with respect to $\Color$ in the subgraph of $G$ induced by $\{v_i, v_{i+1}, \ldots, v_n\}$.
\end{mydefinition}

An ordering $(u_9, u_8, u_7, u_6, u_5, u_4, u_3, u_2, u_1)$ of the graph in Figure~\ref{fig:input-graph-preprocessing} is a colorful degeneracy ordering with respect to $\Color$.

Colorful $s$-cores and a colorful degeneracy ordering can be computed by iteratively removing a vertex with the minimum colorful degree. We use the algorithms proposed in \cite{pan2022fairness, zhang2023fairness} to compute every non-empty colorful $s$-core and a colorful degeneracy ordering, which run in $\mathcal{O}(m)$ time and require $\mathcal{O}(n \cdot |\Color|)$ space.

\subsubsection{Computing a Large Initial Solution}\label{subsec:bounding}
We now introduce our technique to compute a large initial solution for maximum $k$-defective clique search. 

We compute a large $k$-defective clique by iteratively removing a vertex with the smallest colorful degree. 
As the removal progresses, the remaining graph becomes denser and will eventually become a $k$-defective clique. 

\begin{itemize}[leftmargin=1em]
    \item[] 

    \textbf{Initial solution}. Given a graph $G$, a coloring $\chi$ of $G$, and a colorful degeneracy ordering $(v_1, v_2, \ldots, v_n)$ with respect to $\chi$, let $S^*=\{v_i, v_{i+1}, \ldots, v_n\}$ be the longest suffix of $(v_1, v_2, \ldots, v_n)$ such that $S^*$ is a $k$-defective clique in $G$. 
    We use $S^*$ as an initial solution.

\end{itemize}

\begin{myexample}\label{ex:lower-bound}
    Let $k = 1$. For the graph in Figure~\ref{fig:input-graph-preprocessing} and its colorful degeneracy ordering $(u_9, u_8, u_7, u_6, u_5, u_4, u_3, u_2, u_1)$, the longest suffix that forms a 1-defective clique is $\{u_5, u_4, u_3, u_2, u_1\}$; therefore, we use $S^* = \{u_1, u_2, u_3, u_4, u_5\}$ as an initial solution. 
\end{myexample}

Since a colorful degeneracy ordering can be obtained in $\mathcal{O}(m)$ time, an initial solution can be computed in $\mathcal{O}(m)$ time.

\myparagraph{Comparison to techniques of existing algorithms}
Several techniques have been proposed for computing an initial solution~\cite{dai2024theoretically, chang2023efficient, jin2024kd}. However, these techniques are complex and have non-linear time complexities. Specifically, the techniques of \MDC{}~\cite{dai2024theoretically}, \kDC{}~\cite{chang2023efficient}, and \KDClub{}~\cite{jin2024kd} require $\mathcal{O}(n \cdot (\omega_k \cdot \delta + \delta^2))$, $\mathcal{O}(m \cdot \delta)$, and $\mathcal{O}(m \cdot \Delta)$ time, respectively, where $\omega_k$ denotes the size of a maximum $k$-defective clique. In contrast, our technique runs in $\mathcal{O}(m)$ time, which is asymptotically faster than existing algorithms.
  
\subsubsection{Graph Reduction Technique}\label{subsec:reduction}

We now introduce our graph reduction technique to remove vertices and edges that cannot be contained in any maximum $k$-defective clique.

After an initial solution $S^*$ is computed, the threshold parameter $q$ is set to $|S^*| + 1$ to find a larger $k$-defective clique.
Consider a $k$-defective clique $S$ of size at least $q$ in the graph $G$. It is obvious that the degree of every vertex in the induced subgraph $G[S]$ is at least $q - k - 1$; otherwise, $S$ has more than $k$ missing edges. In the following lemma, we show that the colorful degree of every vertex in $G[S]$ must be at least $q - k - 1$. For brevity we abbreviate $d^{\Color}_{G[S \cup \{u\}]}(u)$ as $d^{\Color}_S(u)$.

\begin{mylemma}\label{lemma:color-degree-prune}
    Given a graph $G=(V, E)$, a coloring $\Color$ of $G$, and positive integers $k$ and $q$, let $S$ be a $k$-defective clique of size at least $q$. Then, every vertex $u \in S$ has colorful degree $d^{\Color}_{S}(u) \ge q - k - 1$ in the subgraph induced by $S$.
\end{mylemma}

\begin{proof}
    We only consider the case when $q \ge k + 2$, as the lemma holds trivially otherwise. Suppose not, i.e., there is a vertex $u \in S$ with colorful degree $d^{\Color}_S(u) < q - k - 1$. The set $S$ is then partitioned into two disjoint subsets: $S_1 = \nnbr_S(u)$ and $S_2 = N_S(u)$. We have $|\overline{E}(S_1)| \ge q - d_S(u) - 1$ because $|S| \ge q$, and $|\overline{E}(S_2)| \ge d_S(u) - d^{\Color}_{S}(u)$ since the vertices of the same color must not be adjacent. Therefore, $|\overline{E}(S)| \ge |\overline{E}(S_1)| + |\overline{E}(S_2)| \ge q - 1 - d^{\Color}_{S}(u) > k$ holds. This contradicts the assumption that $d^{\Color}_S(u) < q - k - 1$ holds, thereby proving the lemma.
\end{proof}

By Lemma~\ref{lemma:color-degree-prune}, a vertex with colorful degree less than $q - k - 1$ cannot be contained in any $k$-defective clique of size at least $q$.
Therefore, we compute the reduced graph by iteratively removing a vertex with colorful degree less than $q - k - 1$, while keeping all $k$-defective cliques of size at least $q$. The following lemma formally introduces our reduction technique.

\begin{mylemma}\label{lemma:colorful-core}
Given a graph $G=(V, E)$, a coloring $\Color$ of $G$, and positive integers $k$ and $q$, every $k$-defective clique of size at least $q$ is contained within the colorful $(q - k - 1)$-core of $G$ with respect to $\Color$.
\end{mylemma}

According to Lemma~\ref{lemma:colorful-core}, we reduce the input graph to its colorful $(q - k - 1)$-core.

\begin{myexample}
    In the example graph in Figure~\ref{fig:input-graph-preprocessing}, let $k=1$. When $q = 5$, by Lemma~\ref{lemma:colorful-core}, we reduce the graph to its colorful $3$-core $G[\{u_1, u_2, u_3, u_4, u_5\}]$ by removing vertices $u_6, u_7, u_8$ and $u_9$ from the graph. When $q = 6$, the colorful 4-core is an empty graph.
    Since we obtain an initial solution $S^*$ of size $5$ in Example~\ref{ex:lower-bound}, $q$ is set to $|S^*| + 1 = 6$ and the branch-and-bound search can be avoided.
\end{myexample}

After computing the colorful $(q - k - 1)$-core, we additionally apply the well-known truss-based reduction technique~\cite{gao2022exact, chang2023efficient}, which iteratively removes an edge whose endpoints share at most $q - k - 2$ common neighbors.

Let $g$ be the colorful $(q - k - 1)$-core of $G$, $m_g \le m$ be the number of edges in $g$, and $\delta_g \le \delta$ be the degeneracy of $g$. The time complexity of our reduction technique is $\mathcal{O}(m + m_g \cdot \delta_g)$ because the colorful $(q - k - 1)$ core can be computed in $\mathcal{O}(m)$ time and the truss-based reduction takes $\mathcal{O}(m_g \cdot \delta_g)$ time.

\myparagraph{Comparison to reduction techniques of existing algorithms} 
The core-based reduction technique of \cite{dai2023maximal, dai2024theoretically} reduces the input graph to its $(q - k - 1)$-core.
However, we emphasize that the colorful $(q - k - 1)$-core is always a subgraph of the $(q - k - 1)$-core. For instance, in Figure~\ref{fig:input-graph-preprocessing}, both the 3-core and 4-core of the graph are the graph itself. In contrast, the colorful 3-core is the subgraph induced by $\{u_1, u_2, u_3, u_4, u_5\}$, and the colorful 4-core is an empty graph. Thus, our reduction technique computes a smaller reduced graph than the core-based reduction technique. \KDClub{}~\cite{jin2024kd} also proposes a reduction technique, but it has a high time complexity of $\mathcal{O}(m \cdot \Delta \cdot |\chi|^2)$.

\myparagraph{Applying the reduction technique for maximal $k$-defective clique enumeration}
The graph reduction technique can also be applied for maximal $k$-defective clique enumeration. 
Before the branch-and-bound search, we apply our graph reduction technique using the threshold parameter $q$ to compute the reduced graph. Then, we conduct the branch-and-bound search (Algorithm~\ref{alg:diameter-two}) with the reduced graph. In this way, we can further reduce the search space of our algorithm.

\section{Performance Evaluation}\label{sec:experiments}

\newcommand{\scaleval}{0.75}
\newcommand{\val}{0.18}

In this section, we evaluate the performance of our algorithms as well as competing algorithms for both maximal $k$-defective clique enumeration and maximum $k$-defective clique search. We also evaluate the efficiency of our individual techniques. For brevity, we abbreviate $10^3, 10^6$, and $10^9$ as 1K, 1M, and 1B, respectively.

\subsection{Experimental Setup}\label{subsec:experimental_setup}

\subsubsection{Algorithms}\label{subsec:algorithms}

\myparagraph{Maximal $k$-defective clique enumeration}
We evaluate the efficiency of our algorithm by comparing it against the following two algorithms that solve maximal $k$-defective clique enumeration.
\begin{itemize}
    \item \OursE{}: our algorithm (Algorithm~\ref{alg:diameter-two}, incorporating the graph reduction technique of Section~\ref{subsec:reduction}).
    \item \pivottwo{}: algorithm proposed in \cite{dai2023maximal}.
    \item \pivot{}: state-of-the-art algorithm proposed in \cite{dai2023maximal}, an optimized version of \pivottwo{} enhanced with its pruning technique.
\end{itemize}
We emphasize that \cite{dai2023maximal} is the first paper presenting algorithms solving maximal $k$-defective clique enumeration,
and the two algorithms are by far significantly and consistently faster than the other algorithms presented in \cite{dai2023maximal}. 
In our experiments, all algorithms report only the number of solutions. If the codes were modified to explicitly output all solutions, the total processing time would increase by the time required for outputting.

\myparagraph{Maximum $k$-defective clique search} 
We evaluate the performance of our algorithm by comparing it against the following four state-of-the-art algorithms, which solve the maximum $k$-defective clique search problem and have significantly outperformed other existing algorithms~\cite{gao2022exact, chen2021computing}.
\begin{itemize}
    \item \OursS{}: our algorithm (Algorithm~\ref{alg:maximum}).
    \item \MDC{}: algorithm proposed in \cite{dai2024theoretically}.
    \item \kDCtwo{}: algorithm proposed in \cite{chang2024maximum}, in which its reduction technique \textbf{RR3} is applied.
    \item \kDC{}: algorithm proposed in \cite{chang2023efficient}.
    \item \KDClub{}: algorithm proposed in \cite{jin2024kd}.
\end{itemize}

All the source codes were obtained from the authors of previous studies. All algorithms, including ours, are implemented in C++ and compiled with the \texttt{-O3} optimization flag. The experiments are conducted on a machine running CentOS, equipped with two Intel Xeon E5-2680 v3 2.50GHz CPUs and 256GB of RAM. The source code of this work is available at \url{https://github.com/SNUCSE-CTA/WODC}.

\begin{table}
    \centering
    \caption{Statistics of benchmark graph datasets, where $\omega$ represents the size of the maximum clique in the graph. D1 -- D10 (top) are real-world graph datasets with more than 1M edges, and D11 -- D14 (bottom) are large-scale graph datasets with more than 100M edges.}
    \label{tab:datasets}
    
    \scalebox{0.85}{
    
    \begin{tabular}{c|c|ccccc}
        \toprule
        ID & Dataset & $n$ & $m$ & $\Delta$ & $\delta$ & $\omega$ \\
        \midrule
        D1 &       \textsf{soc-FourSquare}            & 639K       & 3.21M     & 106K & 63    & 30 \\
        D2 &       \textsf{soc-buzznet}         & 101K      & 2.76M       & 64.3K & 153    & 31 \\
        D3 &       \textsf{soc-digg}           & 771K      & 5.91M     & 17.6K & 236    & 50 \\
        D4 &       \textsf{soc-BlogCatalog}       & 88.8K       & 2.09M       & 9,444 & 221    & 45 \\
        D5 &       \textsf{soc-LiveMocha}          & 104K    & 2.19M    & 2,980 & 92    & 15 \\ 
        D6  &      \textsf{socfb-Indiana}     & 29.7K       & 1.31M       & 1,358 & 76    & 48 \\
        D7  &      \textsf{socfb-UF}            & 35.1K       & 1.47M       & 8,246 & 83    & 55 \\
        D8  &      \textsf{socfb-UIllinois}        & 30.8K    & 1.26M    & 4,632 & 85    & 57 \\
        D9 &       \textsf{tech-as-skitter}  & 1.69M  & 11.1M   & 35.5K & 111   & 67 \\
        D10  &      \textsf{web-wikipedia2009}         & 1.86M      & 4.51M    & 2,624 & 66   & 31\\
                \midrule     
        D11 &       \textsf{soc-orkut}         & 3M       & 106M       & 27.5K & 230    & 47 \\
        D12 &       \textsf{enwiki-2023}           & 6.62M      & 150M     & 237K & 188    & 47 \\
        D13 &       \textsf{it-2004}         & 41.3M    & 1.03B    & 1.33M & 3,224   & 3,222 \\
        D14 &       \textsf{sk-2005}            & 50.6M      & 1.81B     & 8.56M & 4,510   & 4,511 \\
        \bottomrule
    \end{tabular}

    }
\end{table}
    
\subsubsection{Datasets}

To evaluate the performance of different algorithms, we use 14 benchmark graph datasets. 
Table~\ref{tab:datasets} shows statistics for the benchmark graph datasets D1 -- D14.
We obtained ten real-world graph datasets\footnote{\url{https://lcs.ios.ac.cn/~caisw/Resource/realworld\%20graphs.tar.gz}} D1 -- D10 with more than 1M edges, which are widely used for evaluating the performance of maximal/maximum $k$-defective clique enumeration/search algorithms~\cite{dai2023maximal, chang2023efficient, jin2024kd, gao2022exact, chen2021computing, dai2024theoretically, chang2024maximum}.
Additionally, we obtained four large-scale graph datasets\footnote{\url{https://law.di.unimi.it/datasets.php}} D11 -- D14 with more than 100M edges to test both the efficiency and scalability of the algorithms.

\subsubsection{Metrics}
We record the total \emph{processing time} required for running an algorithm on a graph dataset for given value(s) of $k$ (and $q$). The processing time recorded reflects the total CPU time, excluding the I/O time spent loading the input graph from disk into main memory. Given that these problems are NP-hard, an algorithm may not terminate within a reasonable time; thus, we set a time limit of 3 hours for the real-world graphs and 24 hours for the large-scale graphs.

Furthermore, we count the number of search tree nodes (i.e., the number of partial solutions) of the algorithms for maximal $k$-defective clique enumeration to verify that the result of our theoretical analysis matches the practical results.

\subsubsection{Parameters}
For maximal $k$-defective clique enumeration, there are two parameters: $k$ and $q$. We select the parameters $k$ and $q$ from $\{1, 3, 5, 7\}$ and $\{10, 20, 30, 40, 50\}$, respectively. 
For large-scale graph datasets with more than 100M edges, we set $q$ to a value slightly smaller than the size of the maximum clique~\cite{dai2022scaling}, because no algorithm can terminate within the time limit for smaller $q$ values.
For maximum $k$-defective clique search, we choose the parameter $k$ from $\{1, 5, 10, 15, 20\}$, as in \cite{chang2023efficient, chen2021computing, jin2024kd, gao2022exact, dai2024theoretically, chang2024maximum}.

\begin{figure*}
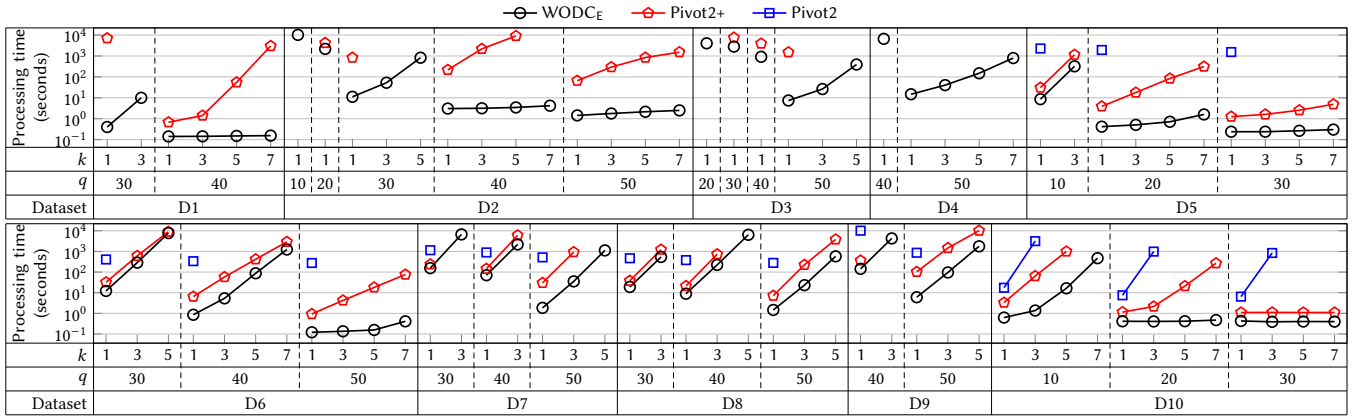

    \centering

    \scalebox{\scaleval}{
          \begin{tikzpicture}
  \begin{axis}[
      hide axis,
      xmin=0, xmax=1,
      ymin=0, ymax=1,
      legend columns=6,
      legend style={draw=none, fill=none},
      height=2cm
  ]
  \addlegendimage{
  line width=0.3mm, color=\OursC, mark=\OursM, mark options={solid, scale=\OursSc}, every mark/.append style={rotate=\OursRo}
  }
  \addlegendentry{\OursE{} \quad\quad}
  
  \addlegendimage{
  line width=0.3mm, color=\PivotC, mark=\PivotM, mark options={solid, scale=\PivotSc}, every mark/.append style={rotate=\PivotRo}
  }
  \addlegendentry{\pivot{} \quad\quad}
  
  \addlegendimage{
  line width=0.3mm, color=\PivottwoC, mark=\PivottwoM, mark options={solid, scale=\PivottwoSc}, every mark/.append style={rotate=\PivottwoRo}
  }
  \addlegendentry{\pivottwo{}}
  \end{axis}
  \end{tikzpicture}
    }

    \vspace*{-1.2mm}
    \hspace*{-4mm}
        \centering
        \scalebox{\scaleval}{\input{Experiments/maximal_1}}
    
    \vspace*{-0.6mm}
    \hspace*{-4mm}
        \scalebox{\scaleval}{\input{Experiments/maximal_2}}

    \caption{Processing time (in seconds) of different algorithms for solving maximal $k$-defective clique enumeration on ten real-world graph datasets. Markers for algorithms that failed to solve the problem within the 3-hour time limit are not shown.}
    \label{fig:maximal-result}
    \illegalvspace{-3mm} %
    \Description{}
\end{figure*}

\begin{figure*}
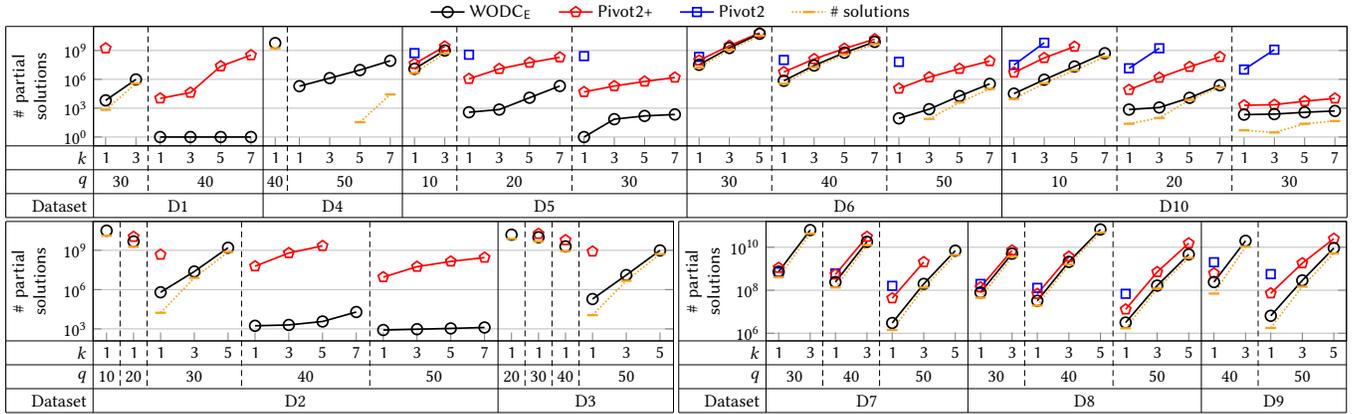

    \centering

    \scalebox{\scaleval}{
          \begin{tikzpicture}
  \begin{axis}[
      hide axis,
      xmin=0, xmax=1,
      ymin=0, ymax=1,
      legend columns=7,
      legend style={draw=none, fill=none},
      height=2cm
  ]
  \addlegendimage{
  line width=0.3mm, color=\OursC, mark=\OursM, mark options={solid, scale=\OursSc}, every mark/.append style={rotate=\OursRo}
  }
  \addlegendentry{\OursE{} \quad\quad}
  
  \addlegendimage{
  line width=0.3mm, color=\PivotC, mark=\PivotM, mark options={solid, scale=\PivotSc}, every mark/.append style={rotate=\PivotRo}
  }
  \addlegendentry{\pivot{} \quad\quad}
  
  \addlegendimage{
  line width=0.3mm, color=\PivottwoC, mark=\PivottwoM, mark options={solid, scale=\PivottwoSc}, every mark/.append style={rotate=\PivottwoRo}
  }
  \addlegendentry{\pivottwo{} \quad\quad}
  
  \addlegendimage{
  line width=0.3mm, color=\SolC, mark=\SolM, densely dotted, mark options={solid, scale=\SolSc, line width=0.4mm}, every mark/.append style={rotate=\SolRo}
  }
  \addlegendentry{\# solutions}
  
  \end{axis}
  \end{tikzpicture}
    }

    \vspace*{-1.2mm}
    \hspace*{-4mm}
        \centering
        \scalebox{\scaleval}{\input{Experiments/maximal_node_1}}

    \vspace*{-0.6mm}
    \hspace*{-4mm}
        \centering
        \scalebox{\scaleval}{\input{Experiments/maximal_node_2}}

    \caption{Number of search tree nodes (\# partial solutions) of different algorithms and number of solutions (\# solutions) on real-world graph datasets. Markers for \# solutions are not shown if there are no solutions.}
    \label{fig:maximal-result-numnodes}
    \illegalvspace{-3mm} %
    \Description{}
\end{figure*}

\begin{figure}
    \centering

    \scalebox{\scaleval}{
          \begin{tikzpicture}
  \begin{axis}[
      hide axis,
      xmin=0, xmax=1,
      ymin=0, ymax=1,
      legend columns=6,
      legend style={draw=none, fill=none},
      height=2cm
  ]
  \addlegendimage{
  line width=0.3mm, color=\OursC, mark=\OursM, mark options={solid, scale=\OursSc}, every mark/.append style={rotate=\OursRo}
  }
  \addlegendentry{\OursE{} \quad\quad}
  
  \addlegendimage{
  line width=0.3mm, color=\PivotC, mark=\PivotM, mark options={solid, scale=\PivotSc}, every mark/.append style={rotate=\PivotRo}
  }
  \addlegendentry{\pivot{} \quad\quad}
  
  \addlegendimage{
  line width=0.3mm, color=\PivottwoC, mark=\PivottwoM, mark options={solid, scale=\PivottwoSc}, every mark/.append style={rotate=\PivottwoRo}
  }
  \addlegendentry{\pivottwo{}}
  \end{axis}
  \end{tikzpicture}
    }

    \vspace*{-1.2mm}
    \hspace*{-4.5mm}
        \centering
        \scalebox{\scaleval}{\pgfplotsset{compat=1.18}

\newcommand{\wwidth}{11.2}
\newcommand{\hheight}{3.7}

\newcommand{\ratioval}{0.83}
\newcommand{\xratio}{0.91}
\newcommand{\yratio}{0.6}

\begin{tikzpicture}

\begin{semilogxaxis}[
    ylabel={Processing time \\ (seconds)},
    ylabel shift=0pt,
    ylabel style={text width=2.1cm, align=center},    xtick={{0,1,2,2.8,3.8,4.8,5.8,6.6,7.6,8.6,9.6,10.4,11.4,12.4,13.4}},
    xticklabels={{1,3,5,1,3,5,7,1,3,5,7,1,3,5,7}},
    ytick={0.001, 0.01, 0.1, 1, 10, 100, 1000, 10000, 100000},
    yticklabels={$10^{-3}$, $10^{-2}$, $10^{-1}$, $10^0$, $10^1$, $10^2$, $10^3$, $10^4$, $10^5$},
    xmode=linear,
    ymode=log,
    xtick pos=bottom,
    log basis y=10,
    yminorgrids=false,
    legend pos=south east,
    yminorgrids=false,
    ymajorgrids=true, 
    width=\wwidth cm,
    height=\hheight cm,
    ymax=220000,
    ymin=5,
    xmin=-0.4,
    xmax=13.8]

  \addplot[line width=0.3mm, color=\PivotC, mark=\PivotM, mark options={solid, scale=\PivotSc}, every mark/.append style={rotate=\PivotRo}] plot coordinates {
    (0,3046.22)(1,40961.83)
  };

  \addplot[line width=0.3mm, color=\OursC, mark=\OursM, mark options={solid, scale=\OursSc}, every mark/.append style={rotate=\OursRo}] plot coordinates {
    (0,69.2311)(1,1078.12)(2,48785.9)
  };

  \addplot[line width=0.3mm, color=\PivotC, mark=\PivotM, mark options={solid, scale=\PivotSc}, every mark/.append style={rotate=\PivotRo}] plot coordinates {
    (2.8,703.95)(3.8,15817.73)
  };

  \addplot[line width=0.3mm, color=\OursC, mark=\OursM, mark options={solid, scale=\OursSc}, every mark/.append style={rotate=\OursRo}] plot coordinates {
    (2.8,23.0809)(3.8,30.999)(4.8,127.687)(5.8,1559.7)
  };

  \addplot[line width=0.3mm, color=\OursC, mark=\OursM, mark options={solid, scale=\OursSc}, every mark/.append style={rotate=\OursRo}] plot coordinates {
    (6.6,1006.3)(7.6,1018.33)(8.6,2136.07)(9.6,30915.5)
  };

  \addplot[line width=0.3mm, color=\OursC, mark=\OursM, mark options={solid, scale=\OursSc}, every mark/.append style={rotate=\OursRo}] plot coordinates {
    (10.4,591.253)(11.4,591.736)(12.4,593.356)(13.4,593.604)
  };

\coordinate (bottomleft) at (rel axis cs:0,0);
\coordinate (bottomright) at (rel axis cs:1,0);
\coordinate (topright) at (rel axis cs:1,1);
\coordinate (topleft) at (rel axis cs:0,1);

\coordinate (x1) at (rel axis cs:-0.16, 0);
\coordinate (x2) at (rel axis cs:-0.16, 1);
\coordinate (x3) at (rel axis cs:-0.16, -0.2);
\coordinate (x4) at (rel axis cs:1, -0.2);
\coordinate (x5) at (rel axis cs:-0.16, -0.4);
\coordinate (x6) at (rel axis cs:1, -0.4);
\coordinate (x7) at (rel axis cs:-0.16, -0.6);
\coordinate (x8) at (rel axis cs:1, -0.6);
\coordinate (x9) at (rel axis cs:0, -0.6);

\coordinate (x10) at (rel axis cs:0, -0.1);
\coordinate (x11) at (rel axis cs:0, -0.3);
\coordinate (x12) at (rel axis cs:0, -0.5);

\coordinate (y0) at (rel axis cs:1.4/14.200000000000001, -0.3);
\coordinate (y1) at (rel axis cs:1.4/14.200000000000001, -0.5);
\coordinate (y2) at (rel axis cs:2.8/14.200000000000001, 1);
\coordinate (y3) at (rel axis cs:2.8/14.200000000000001, -0.6);
\coordinate (y4) at (rel axis cs:4.699999999999999/14.200000000000001, -0.3);
\coordinate (y5) at (rel axis cs:4.699999999999999/14.200000000000001, -0.5);
\coordinate (y6) at (rel axis cs:6.6/14.200000000000001, 1);
\coordinate (y7) at (rel axis cs:6.6/14.200000000000001, -0.6);
\coordinate (y8) at (rel axis cs:8.5/14.200000000000001, -0.3);
\coordinate (y9) at (rel axis cs:8.5/14.200000000000001, -0.5);
\coordinate (y10) at (rel axis cs:10.4/14.200000000000001, 1);
\coordinate (y11) at (rel axis cs:10.4/14.200000000000001, -0.6);
\coordinate (y12) at (rel axis cs:12.3/14.200000000000001, -0.3);
\coordinate (y13) at (rel axis cs:12.3/14.200000000000001, -0.5);
\coordinate (y14) at (rel axis cs:14.200000000000001/14.200000000000001, 1);
\coordinate (y15) at (rel axis cs:14.200000000000001/14.200000000000001, -0.6);

\node at (rel axis cs:8.5/14.2, 0.85) {\texttt{OOM} (\pivot{})};
\node at (rel axis cs:12.3/14.2, 0.85) {\texttt{OOM} (\pivot{})};

\end{semilogxaxis}

\draw (x1) -- (bottomright);
\draw (x2) -- (topright);
\draw (x3) -- (x4);
\draw (x5) -- (x6);
\draw (x7) -- (x8);
\draw (x9) -- (topleft);
\draw (x2) -- (x7);

\node[left] at (x10) {$k$};
\node[left] at (x11) {$q$};
\node[left] at (x12) {Dataset};

\node at (y0) {40};
\node at (y1) {D11};
\draw (y2) -- (y3);
\node at (y4) {40};
\node at (y5) {D12};
\draw (y6) -- (y7);
\node at (y8) {3000};
\node at (y9) {D13};
\draw (y10) -- (y11);
\node at (y12) {4000};
\node at (y13) {D14};
\draw (y14) -- (y15);

\end{tikzpicture}}

    \caption{Processing time (in seconds) of different algorithms for solving maximal $k$-defective clique enumeration on four large-scale graph datasets. \pivottwo{} failed to solve the problem within the 24-hour time limit for all large-scale datasets. \texttt{OOM} indicates ``out of memory''; \pivot{} runs out of memory when solving graphs with more than 1B edges (D13 and D14).}
    \label{fig:maximal-result-large}
    \Description{}
\end{figure}
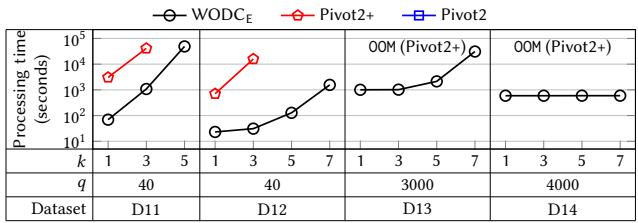

\begin{figure*}
    \centering

    \scalebox{\scaleval}{
          \begin{tikzpicture}
  \begin{axis}[
      hide axis,
      xmin=0, xmax=1,
      ymin=0, ymax=1,
      legend columns=5,
      legend style={draw=none, fill=none},
      height=2cm
  ]
  \addlegendimage{
  line width=0.3mm, color=\OursC, mark=\OursM, mark options={solid, scale=\OursSc}, every mark/.append style={rotate=\OursRo}
  }
  \addlegendentry{\OursS{} \quad\quad}
  
  \addlegendimage{
  line width=0.3mm, color=\MDCC, mark=\MDCM, mark options={solid, scale=\MDCSc}, every mark/.append style={rotate=\MDCRo}
  }
  \addlegendentry{\MDC{} \quad\quad}
  
  \addlegendimage{
  line width=0.3mm, color=\kDCtwoC, mark=\kDCtwoM, mark options={solid, scale=\kDCtwoSc}, every mark/.append style={rotate=\kDCtwoRo}
  }
  \addlegendentry{\kDCtwo{} \quad\quad}
  
  \addlegendimage{
  line width=0.3mm, color=\kDCC, mark=\kDCM, mark options={solid, scale=\kDCSc}, every mark/.append style={rotate=\kDCRo}
  }
  \addlegendentry{\kDC{} \quad\quad}

  \addlegendimage{
  line width=0.3mm, color=\KDClubC, mark=\KDClubM, mark options={solid, scale=\KDClubSc}, every mark/.append style={rotate=\KDClubRo}
  }
  \addlegendentry{\KDClub{}}
  
  \end{axis}
  \end{tikzpicture}
    }

    \vspace*{-1.2mm}
    \hspace*{-4mm}
        \centering
        \scalebox{\scaleval}{\input{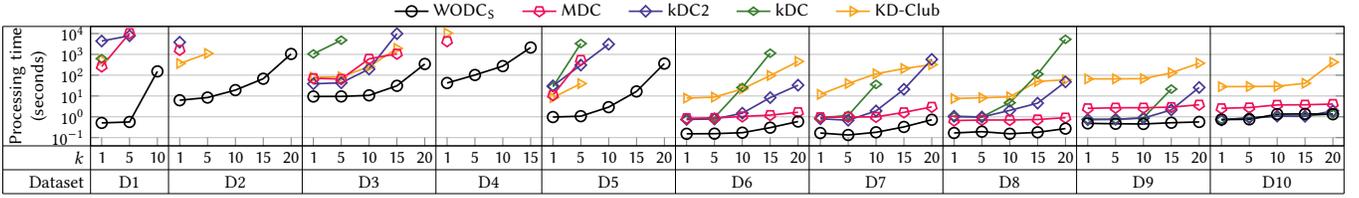}}

    \caption{Processing time (in seconds) of different algorithms for solving maximum $k$-defective clique search on ten real-world graphs. Markers for algorithms that failed to solve the problem within the 3-hour time limit are not shown.}
    \label{fig:maximum-result}
    \illegalvspace{-3mm} %
    \Description{}
\end{figure*}

\begin{figure}
    \centering

    \scalebox{\scaleval}{
          \begin{tikzpicture}
  \begin{axis}[
      hide axis,
      xmin=0, xmax=1,
      ymin=0, ymax=1,
      legend columns=5,
      legend style={draw=none, fill=none},
      height=2cm
  ]
  \addlegendimage{
  line width=0.3mm, color=\OursC, mark=\OursM, mark options={solid, scale=\OursSc}, every mark/.append style={rotate=\OursRo}
  }
  \addlegendentry{\OursS{} \quad\quad}
  
  \addlegendimage{
  line width=0.3mm, color=\MDCC, mark=\MDCM, mark options={solid, scale=\MDCSc}, every mark/.append style={rotate=\MDCRo}
  }
  \addlegendentry{\MDC{} \quad\quad}
  
  \addlegendimage{
  line width=0.3mm, color=\kDCtwoC, mark=\kDCtwoM, mark options={solid, scale=\kDCtwoSc}, every mark/.append style={rotate=\kDCtwoRo}
  }
  \addlegendentry{\kDCtwo{} \quad\quad}
  
  \addlegendimage{
  line width=0.3mm, color=\kDCC, mark=\kDCM, mark options={solid, scale=\kDCSc}, every mark/.append style={rotate=\kDCRo}
  }
  \addlegendentry{\kDC{} \quad\quad}

  \addlegendimage{
  line width=0.3mm, color=\KDClubC, mark=\KDClubM, mark options={solid, scale=\KDClubSc}, every mark/.append style={rotate=\KDClubRo}
  }
  \addlegendentry{\KDClub{}}
  
  \end{axis}
  \end{tikzpicture}
    }

    \vspace*{-1.2mm}
    \hspace*{-4mm}
        \centering
        \scalebox{\scaleval}{\pgfplotsset{compat=1.18}

\newcommand{\wwidth}{11.2}
\newcommand{\hheight}{3.7}

\newcommand{\ratioval}{0.83}
\newcommand{\xratio}{0.91}
\newcommand{\yratio}{0.6}

\begin{tikzpicture}

\begin{semilogxaxis}[
    ylabel={Processing time \\ (seconds)},
    ylabel shift=0pt,
    ylabel style={text width=2.1cm, align=center},
    xtick={{0,1,2,3,4,4.8,5.8,6.8,7.8,8.8,9.600000000000001,10.600000000000001,11.600000000000001,12.600000000000001,13.600000000000001,14.400000000000002,15.400000000000002,16.400000000000002,17.400000000000002,18.400000000000002}},
    xticklabels={{1,5,10,15,20,1,5,10,15,20,1,5,10,15,20,1,5,10,15,20}},
    ytick={0.001, 0.01, 0.1, 1, 10, 100, 1000, 10000, 100000},
    yticklabels={$10^{-3}$, $10^{-2}$, $10^{-1}$, $10^0$, $10^1$, $10^2$, $10^3$, $10^4$, $10^5$},
    xmode=linear,
    ymode=log,
    xtick pos=bottom,
    log basis y=10,
    yminorgrids=false,
    legend pos=south east,
    yminorgrids=false,
    ymajorgrids=true, 
    width=\wwidth cm,
    height=\hheight cm,
    ymax=220000,
    ymin=5,
    xmin=-0.4,
    xmax=18.800000000000004]

  \addplot[line width=0.3mm, color=\KDClubC, mark=\KDClubM, mark options={solid, scale=\KDClubSc}, every mark/.append style={rotate=\KDClubRo}] plot coordinates {
    (0,1706.82)(1,2238.81)(2,9416.13)
  };

  \addplot[line width=0.3mm, color=\kDCC, mark=\kDCM, mark options={solid, scale=\kDCSc}, every mark/.append style={rotate=\kDCRo}] plot coordinates {
    (0,420.642505)(1,14105.060337)
  };

  \addplot[line width=0.3mm, color=\kDCtwoC, mark=\kDCtwoM, mark options={solid, scale=\kDCtwoSc}, every mark/.append style={rotate=\kDCtwoRo}] plot coordinates {
    (0,97.191406)(1,177.303928)(2,666.202893)(3,9346.23493)
  };

  \addplot[line width=0.3mm, color=\MDCC, mark=\MDCM, mark options={solid, scale=\MDCSc}, every mark/.append style={rotate=\MDCRo}] plot coordinates {
    (0,143.806)(1,186.693)(2,268.453)(3,52442.233)(4,54426.583)
  };

  \addplot[line width=0.3mm, color=\OursC, mark=\OursM, mark options={solid, scale=\OursSc}, every mark/.append style={rotate=\OursRo}] plot coordinates {
    (0,122.615)(1,130.844)(2,148.043)(3,173.029)(4,220.133)
  };

  \addplot[line width=0.3mm, color=\KDClubC, mark=\KDClubM, mark options={solid, scale=\KDClubSc}, every mark/.append style={rotate=\KDClubRo}] plot coordinates {
    (4.8,2893.27)(5.8,3816.43)(6.8,6734.37)(7.8,20010.8)
  };

  \addplot[line width=0.3mm, color=\kDCC, mark=\kDCM, mark options={solid, scale=\kDCSc}, every mark/.append style={rotate=\kDCRo}] plot coordinates {
    (4.8,68.416116)(5.8,84.442925)(6.8,135.155672)(7.8,1327.883622)(8.8,74515.410087)
  };

  \addplot[line width=0.3mm, color=\kDCtwoC, mark=\kDCtwoM, mark options={solid, scale=\kDCtwoSc}, every mark/.append style={rotate=\kDCtwoRo}] plot coordinates {
    (4.8,70.874003)(5.8,88.447653)(6.8,92.025399)(7.8,110.96184)(8.8,536.85977)
  };

  \addplot[line width=0.3mm, color=\MDCC, mark=\MDCM, mark options={solid, scale=\MDCSc}, every mark/.append style={rotate=\MDCRo}] plot coordinates {
    (4.8,122.927)(5.8,121.422)(6.8,136.217)(7.8,159.941)(8.8,171.018)
  };

  \addplot[line width=0.3mm, color=\OursC, mark=\OursM, mark options={solid, scale=\OursSc}, every mark/.append style={rotate=\OursRo}] plot coordinates {
    (4.8,25.5653)(5.8,25.8925)(6.8,27.301)(7.8,29.2915)(8.8,38.7201)
  };

  \addplot[line width=0.3mm, color=\kDCC, mark=\kDCM, mark options={solid, scale=\kDCSc}, every mark/.append style={rotate=\kDCRo}] plot coordinates {
    (9.600000000000001,261.511085)(10.600000000000001,342.228891)(11.600000000000001,367.131035)(12.600000000000001,370.806385)(13.600000000000001,372.118832)
  };

  \addplot[line width=0.3mm, color=\kDCtwoC, mark=\kDCtwoM, mark options={solid, scale=\kDCtwoSc}, every mark/.append style={rotate=\kDCtwoRo}] plot coordinates {
    (9.600000000000001,221.929366)(10.600000000000001,300.852086)(11.600000000000001,356.61525)(12.600000000000001,366.478238)(13.600000000000001,303.329141)
  };

  \addplot[line width=0.3mm, color=\MDCC, mark=\MDCM, mark options={solid, scale=\MDCSc}, every mark/.append style={rotate=\MDCRo}] plot coordinates {
    (9.600000000000001,163.075)(10.600000000000001,21133.9)(11.600000000000001,27839.48)(12.600000000000001,28167.824)(13.600000000000001,28528.398)
  };

  \addplot[line width=0.3mm, color=\OursC, mark=\OursM, mark options={solid, scale=\OursSc}, every mark/.append style={rotate=\OursRo}] plot coordinates {
    (9.600000000000001,18.4059)(10.600000000000001,225.706)(11.600000000000001,306.091)(12.600000000000001,402.798)(13.600000000000001,440.847)
  };

  \addplot[line width=0.3mm, color=\kDCC, mark=\kDCM, mark options={solid, scale=\kDCSc}, every mark/.append style={rotate=\kDCRo}] plot coordinates {
    (14.400000000000002,30.15925)(15.400000000000002,415.196356)(16.400000000000002,418.155192)(17.400000000000002,66.500506)(18.400000000000002,66.557092)
  };

  \addplot[line width=0.3mm, color=\kDCtwoC, mark=\kDCtwoM, mark options={solid, scale=\kDCtwoSc}, every mark/.append style={rotate=\kDCtwoRo}] plot coordinates {
    (14.400000000000002,29.997462)(15.400000000000002,83.173507)(16.400000000000002,67.134323)(17.400000000000002,66.662547)(18.400000000000002,83.095237)
  };

  \addplot[line width=0.3mm, color=\MDCC, mark=\MDCM, mark options={solid, scale=\MDCSc}, every mark/.append style={rotate=\MDCRo}] plot coordinates {
    (14.400000000000002,150.664)(15.400000000000002,207.735)(16.400000000000002,205.725)(17.400000000000002,204.032)(18.400000000000002,196.888)
  };

  \addplot[line width=0.3mm, color=\OursC, mark=\OursM, mark options={solid, scale=\OursSc}, every mark/.append style={rotate=\OursRo}] plot coordinates {
    (14.400000000000002,39.8437)(15.400000000000002,107.46)(16.400000000000002,81.6025)(17.400000000000002,80.6918)(18.400000000000002,80.7713)
  };

\coordinate (bottomleft) at (rel axis cs:0,0);
\coordinate (bottomright) at (rel axis cs:1,0);
\coordinate (topright) at (rel axis cs:1,1);
\coordinate (topleft) at (rel axis cs:0,1);

\coordinate (x1) at (rel axis cs:-0.16, 0);
\coordinate (x2) at (rel axis cs:-0.16, 1);
\coordinate (x3) at (rel axis cs:-0.16, -0.2);
\coordinate (x4) at (rel axis cs:1, -0.2);
\coordinate (x5) at (rel axis cs:-0.16, -0.4);
\coordinate (x6) at (rel axis cs:1, -0.4);
\coordinate (x7) at (rel axis cs:0, -0.4);

\coordinate (x10) at (rel axis cs:0, -0.1);
\coordinate (x11) at (rel axis cs:0, -0.3);

\coordinate (y0) at (rel axis cs:2.4/19.200000000000003, -0.3);
\coordinate (y1) at (rel axis cs:4.8/19.200000000000003, 1);
\coordinate (y2) at (rel axis cs:4.8/19.200000000000003, -0.4);
\coordinate (y3) at (rel axis cs:7.200000000000001/19.200000000000003, -0.3);
\coordinate (y4) at (rel axis cs:9.600000000000001/19.200000000000003, 1);
\coordinate (y5) at (rel axis cs:9.600000000000001/19.200000000000003, -0.4);
\coordinate (y6) at (rel axis cs:12.000000000000002/19.200000000000003, -0.3);
\coordinate (y7) at (rel axis cs:14.400000000000002/19.200000000000003, 1);
\coordinate (y8) at (rel axis cs:14.400000000000002/19.200000000000003, -0.4);
\coordinate (y9) at (rel axis cs:16.800000000000004/19.200000000000003, -0.3);
\coordinate (y10) at (rel axis cs:19.200000000000003/19.200000000000003, 1);
\coordinate (y11) at (rel axis cs:19.200000000000003/19.200000000000003, -0.4);

\end{semilogxaxis}

\draw (x1) -- (bottomright);
\draw (x2) -- (topright);
\draw (x3) -- (x4);
\draw (x5) -- (x6);
\draw (x7) -- (topleft);
\draw (x2) -- (x5);

\node[left] at (x10) {$k$};
\node[left] at (x11) {Dataset};

\node at (y0) {D11};
\draw (y1) -- (y2);
\node at (y3) {D12};
\draw (y4) -- (y5);
\node at (y6) {D13};
\draw (y7) -- (y8);
\node at (y9) {D14};
\draw (y10) -- (y11);

\end{tikzpicture}}

    \caption{Processing time (in seconds) of different algorithms for solving maximum $k$-defective clique search on four large-scale graphs. Markers for algorithms that failed to solve the problem within the 24-hour time limit are not shown.}
    \label{fig:maximum-result-large}
    \Description{}
\end{figure}
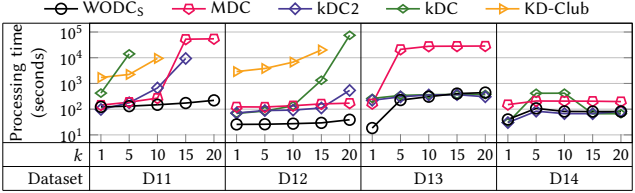

\begin{figure*}
    \centering

    \scalebox{\scaleval}{
          \begin{tikzpicture}
  \begin{axis}[
      hide axis,
      xmin=0, xmax=1,
      ymin=0, ymax=1,
      legend columns=6,
      legend style={draw=none, fill=none},
      height=2cm
  ]
  \addlegendimage{
  line width=0.3mm, color=\OursC, mark=\OursM, mark options={solid, scale=\OursSc}, every mark/.append style={rotate=\OursRo}
  }
  \addlegendentry{\OursE{} \quad\quad}
  
  \addlegendimage{
  line width=0.3mm, color=\OurspC, mark=\OurspM, mark options={solid, scale=\OurspSc}, every mark/.append style={rotate=\OurspRo}
  }
  \addlegendentry{\OursE{}\textsf{-pivot} \quad\quad}
  
  \addlegendimage{
  line width=0.3mm, color=\OurssC, mark=\OurssM, mark options={solid, scale=\OurssSc}, every mark/.append style={rotate=\OurssRo}
  }
  \addlegendentry{\OursE{}\textsf{-selection} \quad\quad}
  
  \addlegendimage{
  line width=0.3mm, color=\OursrC, mark=\OursrM, mark options={solid, scale=\OursrSc}, every mark/.append style={rotate=\OursrRo}
  }
  \addlegendentry{\OursE{}\textsf{-reduction}}
  
  \end{axis}
  \end{tikzpicture}
    }

    \vspace*{-1.2mm}
    \hspace*{-4mm}
        \centering
        \scalebox{\scaleval}{\input{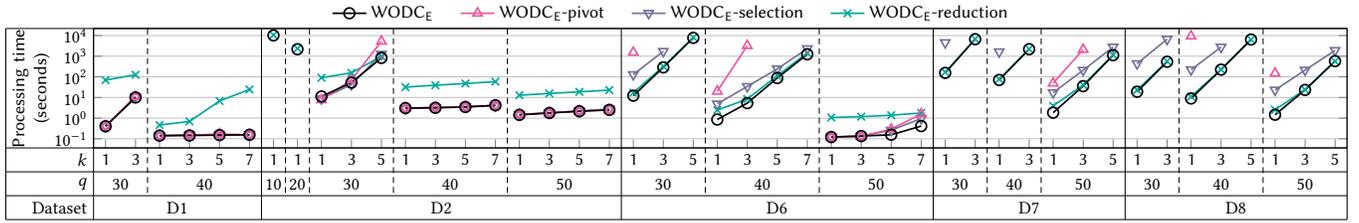}}

    \caption{Processing time (in seconds) of \OursE{} and its variants. Markers for algorithms that failed to solve the problem within the 3-hour time limit are not shown.}
    \label{fig:maximal-abl}
    \illegalvspace{-3mm} %
    \Description{}
\end{figure*}

\subsection{Maximal \texorpdfstring{$k$}{k}-Defective Clique Enumeration}
In this subsection, we evaluate the effectiveness of our algorithm \OursE{} against the existing state-of-the-art algorithms.

\myparagraph{Efficiency of different algorithms}
Figure~\ref{fig:maximal-result} and Figure~\ref{fig:maximal-result-numnodes} show the processing time and the number of search tree nodes of different algorithms on ten real-world graph datasets, respectively. Instances which are not displayed in the figures were unsolved by all algorithms within the time limit, e.g., when $k \ge 3$ and $q \le 40$ in D3 of Figure~\ref{fig:maximal-result}.
We can observe from Figure~\ref{fig:maximal-result} that \OursE{} is significantly and consistently faster for all datasets and all parameters.
Notably, when $k=1$ and $q=30$ on D1, our algorithm achieves $16,290 \times$ speedup over \pivot{}, and \pivottwo{} fails to solve the problem within the time limit. This is because the search space of our algorithm is not only theoretically worst-case optimal but also significantly smaller in practice; on D1 when $k=1$ and $q=30$, our algorithm generates only 6.5K search tree nodes, compared to 1.71B nodes generated by \pivot{} (as shown in Figure~\ref{fig:maximal-result-numnodes}). We also observe that the search space of our algorithm remains remarkably small on other datasets---for instance, on D2 with $q = 40$ and $50$, it is four to six orders of magnitude smaller than that of \pivot{}, and on D5 with $q = 20$ and $30$, it is three to five orders of magnitude smaller. These reductions in search space result in up to several thousand times faster processing time on D2 and D5, as shown in Figure~\ref{fig:maximal-result}.
Moreover, in many cases, only \OursE{} solves the problem within the time limit (e.g., on D1 when $k = 3$ and $q = 30$, and on D4).
In most cases, \pivottwo{} fails to solve the problem within the time limit.
In general, the speedup of \OursE{} over \pivot{} increases as $k$ grows (e.g., on D1, D2, and D5). This is because the search space size of \pivot{} approaches $\mathcal{O}(2^n)$ as $k$ grows; in contrast, our algorithm does not exhibit such growth.
Similarly, as the size constraint $q$ grows (e.g., on D3 when $k=1$), the speedup improves, since our framework finds large solutions more efficiently than \pivot{}.

\myparagraph{Evaluation of the search space size}
Figure~\ref{fig:maximal-result-numnodes} shows the number of solutions, as well as the number of partial solutions generated by different algorithms.
The search space of our algorithm is \mbox{consistently} smaller than that of competing algorithms for all datasets and all \mbox{parameters}. This result matches the theoretical analysis, as the number of search tree nodes is asymptotically smaller than the competing algorithms (see Table~\ref{tab:search-space}). Moreover, in most cases, the number of search tree nodes in our algorithm closely matches the number of solutions. This is because the search space of \OursE{} is worst-case optimal. In some cases, \OursE{} generates only a single partial solution (e.g., on D1 when $q = 40$ and on D5 when $k = 1$ and $q = 30$) because our reduction technique reduces the input graph to an empty graph. 
We also observe that our clique-first approach effectively mitigates the combinatorial explosion of small partial solutions.
For example, on D10 with $k=7$ and $q=20$, \pivot{} generates 215M partial solutions, 96\% of which are small (i.e., of size at most 5). In contrast, our algorithm generates only 242K partial solutions.

\myparagraph{Scalability}
We also evaluate the scalability of our algorithm by running the algorithms on large-scale graphs with more than 100M edges; see Figure~\ref{fig:maximal-result-large}. \OursE{} always outperforms competing algorithms in terms of processing time across all datasets and all parameters. \pivottwo{} fails to solve the problem within the 24-hour time limit for all datasets. For two datasets, D13 and D14, with more than 1B edges, our algorithm is able to enumerate all solutions within the time limit. However, \pivot{} runs out of memory; these results show that our algorithm is more scalable than the competitors.

\myparagraph{Computational challenge of finding defective cliques}
Maximal \emph{clique} listing may be a problem which is NP-hard, but "easy" in most real-world graphs. We conducted an experiment to compare maximal clique listing against maximal defective clique listing. The well-known maximal clique listing algorithm by Eppstein et al.~\cite{eppstein2013listing} successfully solved all instances (except D4) within 30 minutes on datasets D1--D10. In contrast, when $k = 5$ and $q=10$ (and other cases) in Figure~\ref{fig:maximal-result}, none of the maximal $k$-defective clique enumeration algorithms were able to solve instances on D1--D9 within the 3-hour time limit. This shows the greater computational challenge posed by the defective clique problem.

\subsection{Maximum \texorpdfstring{$k$}{k}-Defective Clique Search}
In this subsection, we evaluate the effectiveness of our algorithm \OursS{} against the existing state-of-the-art algorithms.

\myparagraph{Efficiency of different algorithms}
Figure~\ref{fig:maximum-result} shows the processing time on ten real-world graph datasets with varying $k$. Instances where all algorithms exceed the time limit are not shown, e.g., when $k \ge 15$ in D1.
\OursS{} is significantly faster for most of the datasets and varying $k$. Notably, when $k=5$ on D1, our algorithm achieves a speedup of $18,339 \times$ and $13,869 \times$ over \MDC{} and \kDCtwo{}, respectively (other competing algorithms do not terminate within the time limit). 
This is because our clique-first approach quickly finds larger \mbox{$k$-defective} cliques (and updates $q$ accordingly) by avoiding the combinatorial explosion of partial solutions; 
on D1 when $k=5$, the number of partial solutions generated by \MDC{} and \kDCtwo{} are 48.6M and 30K, respectively. In contrast, our algorithm generates only 6.51K partial solutions in total. 
Additionally, when $k=1$ or $5$ on D2 and D5, the search space of our algorithm is one to three orders of magnitude smaller than those of \MDC{}, \kDCtwo{}, and \kDC{}. As a result, our algorithm achieves a speedup of one to four orders of magnitude compared to competing algorithms on D2 and D5. 
Though the search space of \KDClub{} is smaller than \MDC{} or \kDCtwo{}, \KDClub{} is always slower than \OursS{} due to its expensive reduction technique.
Also, \OursS{} computes a large initial solution, which sets the threshold parameter $q$ to a high value before the branch-and-bound search begins. For example, on D6--D8 when $k \le 10$, our algorithm finds a maximum $k$-defective clique as the initial solution, thereby setting $q$ to its maximum possible value. As a result, our algorithm completes within a second on D6--D8 for all values of $k$. In general, the speedup of \OursS{} over the competing algorithms increases as $k$ grows. This is because the search space size of the competing algorithms approaches $\mathcal{O}(2^n)$ as $k$ grows.
Furthermore, as $k$ increases, there are many cases where our algorithm solves the problem within the time limit, while all competing algorithms fail (e.g., on D1–D5). 
This demonstrates the efficiency of our algorithm for large values of $k$ compared to the competing algorithms.

\myparagraph{Scalability}
We also evaluate the scalability of our algorithm with large-scale graph datasets with more than 100M edges; see Figure~\ref{fig:maximum-result-large}. In most cases, our algorithm is faster than all other competing algorithms. Notably, on D11 when $k = 20$, our algorithm solves the problem within four minutes, whereas all other competing algorithms, except \MDC{}, fail to solve it within the 24-hour time limit (\MDC{} takes more than 15 hours). On D13 and D14, \KDClub{} fails to solve the problem within the 24-hour time limit for all $k$-values. Across all datasets and all $k$ values, our algorithm completes within ten minutes; these results indicate that \OursS{} is scalable for large-scale graphs.

\subsection{Effectiveness of Individual Techniques}\label{subsec:ind-techniques}

In this subsection, we evaluate the effectiveness of our individual techniques for maximal $k$-defective clique enumeration and maximum $k$-defective clique search.

\myparagraph{Maximal $k$-defective clique enumeration}
We implemented the following variants of \OursE{} to evaluate the effectiveness of individual techniques.

\begin{itemize}
    \item \OursE{}\textsf{-pivot}: \OursE{} without the pivoting technique in Section~\ref{subsec:pivot}.
    \item \OursE{}\textsf{-selection}: \OursE{} in which the pivot vertex is randomly selected (i.e., without our pivot selection strategy).
    \item \OursE{}\textsf{-reduction}: \OursE{} without the reduction technique in Section~\ref{subsec:reduction}.
\end{itemize}

Figure~\ref{fig:maximal-abl} demonstrates the effectiveness of our individual techniques on datasets that show representative cases.
\begin{itemize}
    \item Our pivoting technique plays a crucial role in reducing the search space, especially for small $q$ values and dense input graphs (e.g., D6--D8). In many cases, \OursE{} solves the problem within the time limit, whereas \OursE{}\textsf{-pivot} fails to do so (on D6--D8). This result highlights the effectiveness of the pivoting technique in eliminating unnecessary branches and reducing the search space.
    \item Our pivot selection strategy significantly reduces the search space. Specifically, \OursE{} runs one to two orders of magnitude faster than \OursE{}\textsf{-selection} on D6--D8 when $k=1$ and $q=30$. This demonstrates that our pivot selection strategy not only leads to a worst-case optimal search space theoretically but also effectively reduces the search space in practice.
    \item A comparison between \OursE{} and \OursE{}\textsf{-reduction} demonstrates that our reduction technique significantly reduces processing time, particularly for large $q$ values and sparse input graphs (e.g., on D1 and D2). This is because as $q$ increases, more vertices and edges are pruned by our reduction technique. On D1 when $q = 40$, our reduction technique reduces the input graph to an empty graph. These results confirm that our reduction technique substantially reduces the search space.
\end{itemize}

\myparagraph{Maximum $k$-defective clique search}
Table~\ref{tab:table-preproc} presents the results of the practical techniques (initial solution and graph reduction) used in \OursS{}, \MDC{}, \kDC{}, and \KDClub{} (the results of \kDCtwo{} are omitted because it uses the same technique as \kDC{}), with time measured for executing Lines~\ref{line4:setq} and~\ref{line4:reduce} in Algorithm~\ref{alg:maximum}.
Our algorithm is the fastest on all datasets except D2. This is due to the lower time complexity of our practical techniques compared to competitors.
When comparing initial solutions, on D5--D10, our initial solution was at least as large as those computed by other algorithms. 
Regarding the reduced graph size, our algorithm produces the smallest reduced graph in the shortest time for most datasets (D1, D6--D10). Notably, even when the initial solution size is smaller, our reduction technique still achieves the smallest reduced graph on D1, demonstrating its effectiveness.

\begin{table}[t]
    \setlength{\tabcolsep}{1.8pt}
    \centering
    \caption{Time for computing the initial solution and reduced graph (in seconds), size of the initial solution ($|S^*|$), and percentage of remaining edges in the reduced graph ($\% E_g$) for $k=5$ on ten real-world graph datasets.}
    \label{table:results}
    \scalebox{0.85}{
    
\begin{tabular}{c|ccc|ccc|ccc|ccc}
    \toprule
    \multirow{2}{*}{Dataset} & \multicolumn{3}{c|}{\OursS{}} & \multicolumn{3}{c|}{\MDC{}} & \multicolumn{3}{c|}{\kDC{}} & \multicolumn{3}{c}{\KDClub} \\
    & Time & $|S^*|$ & $\% E_g$ & Time & $|S^*|$ & $\% E_g$ & Time & $|S^*|$ & $\% E_g$ & Time & $|S^*|$ & $\% E_g$ \\
    \midrule
D1 & 0.23 & 32 & 7.90 & 1.07 & 33 & 28.22 & 1.22 & 34 & 13.45 & 6084 & 30 & 7.97 \\
D2 & 3.05 & 33 & 15.49 & 3.02 & 33 & 78.06 & 2.98 & 35 & 13.36 & 813 & 29 & 1.77 \\
D3 & 6.41 & 43 & 14.81 & 7.85 & 53 & 50.75 & 16.0 & 55 & 6.19 & 33.5 & 49 & 0.34 \\
D4 & 2.61 & 48 & 31.46 & 3.74 & 46 & 73.89 & 4.98 & 49 & 30.48 & 9700 & 44 & 7.06 \\
D5 & 0.60 & 18 & 1.59 & 2.12 & 17 & 88.61 & 1.09 & 18 & 1.59 & 21.6 & 15 & 0.67 \\
D6 & 0.10 & 51 & 0.16 & 0.51 & 51 & 75.53 & 0.72 & 51 & 1.39 & 4.08 & 48 & 0.72 \\
D7 & 0.09 & 59 & 0.22 & 0.51 & 59 & 57.35 & 0.67 & 59 & 3.60 & 31.4 & 55 & 1.51 \\
D8 & 0.11 & 62 & 0 & 0.45 & 62 & 52.80 & 1.00 & 62 & 0.78 & 3.60 & 57 & 0.42 \\
D9 & 0.44 & 70 & 0.03 & 0.89 & 70 & 2.95 & 0.76 & 70 & 0.15 & 2.22 & 67 & 0.04 \\
D10 & 0.58 & 32 & 0.03 & 1.23 & 31 & 1.50 & 0.91 & 31 & 0.04 & 0.92 & 31 & 0.04 \\
    \bottomrule
\end{tabular}

    \label{tab:table-preproc}
    }
\end{table}

\subsection{Case Study}
In this subsection, we present a use case study to demonstrate the real-world applicability of our approach.

\subsubsection{Comparison of Relaxed Cliques}
We first give a conceptual comparison of relaxed cliques based on their definitions.
The $k$-plex is defined by degrees of vertices~\cite{conte2018d2k, chang2024maximumplex, dai2022scaling}, and the quasi-clique is defined by the density of a graph~\cite{guo2020scalable, khalil2022parallel, yu2023fast}.
However, the defective clique specifies directly the number of missing edges.
More specifically, the $k$-plex and the quasi-clique are defined as follows:
\begin{itemize}
\item $k$-plex: Each vertex is allowed to have at most $k-1$ missing edges.
\item $\gamma$-quasi-clique: Each vertex is connected to at least a $\gamma$ fraction of the other vertices.
\end{itemize}
Note that both the 1-plex and the 1-quasi-clique are equivalent to a clique.
Figure~\ref{fig:models} illustrates examples of relaxed cliques with six vertices.
A 2-plex of a clique with $n$ vertices allows at most $\lfloor \frac{n}{2} \rfloor$ missing edges, as shown in Figure~\ref{fig:model2}.
Similarly, a $\gamma$-quasi-clique with $n$ vertices allows up to $\lfloor (1-\gamma) \cdot {n \choose 2} \rfloor$ missing edges, e.g., a \mbox{0.8-quasi-clique} with six vertices (shown in Figure~\ref{fig:model1}) allows up to 3 missing edges. That is, there is no way of specifying the case of just one missing edge (which is 1-defective clique) in the $k$-plex and $\gamma$-quasi-clique models. Therefore, the defective clique is arguably the most intuitive, and it is a better metric in specifying the relaxedness of a clique. In the following, we will show that the defective clique is also more effective than the $k$-plex and the quasi-clique in a downstream task.

\subsubsection{Use Case Study}
To show that the defective clique model achieves higher accuracy than other models in a downstream task (i.e., link prediction),
we performed link prediction on PPI network using  relaxed clique models. To enumerate maximal relaxed cliques, we employed state-of-the-art enumeration algorithms~\cite{dai2022scaling, yu2023fast}.

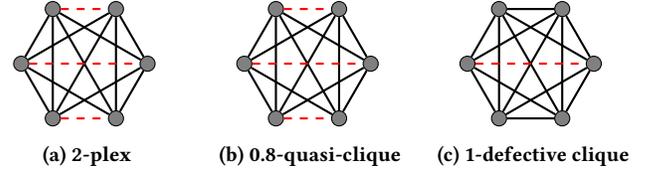
\begin{figure}
    \centering
    \begin{subfigure}{0.3\linewidth}
        \centering
        \newcommand{\x}{0.7}
\newcommand{\y}{0.6}

\newcommand{\scll}{1.1}
\newcommand{\scl}{1.25}

\newcommand{\xx}{5.8}
\newcommand{\yy}{-1}

\hspace{-9mm}

\begin{tikzpicture}  
  [scale=.6,auto=center,every node/.style={circle, inner sep=0.0mm, draw, minimum width=2mm, fill=gray}, > = stealth]

  \node [] (u1) at (-1*\x, 0*\x) {};
  \node [] (u2) at (1*\x, 0*\x) {};
  \node [] (u3) at (-2*\x, 1.732*\x) {};
  \node [] (u4) at (2*\x, 1.732*\x) {};
  \node [] (u5) at (-1*\x, 1.732*2*\x) {};
  \node [] (u6) at (1*\x, 1.732*2*\x) {};

  \path[-, thick, dashed, red] (u1) edge (u2); 
  \path[-, thick] (u1) edge (u3); 
  \path[-, thick] (u1) edge (u4); 
  \path[-, thick] (u1) edge (u5); 
  \path[-, thick] (u1) edge (u6); 
  \path[-, thick] (u2) edge (u3); 
  \path[-, thick] (u2) edge (u4); 
  \path[-, thick] (u2) edge (u5); 
  \path[-, thick] (u2) edge (u6); 
  \path[-, thick, dashed, red] (u3) edge (u4); 
  \path[-, thick] (u3) edge (u5); 
  \path[-, thick] (u3) edge (u6); 
  \path[-, thick] (u4) edge (u5); 
  \path[-, thick] (u4) edge (u6); 
  \path[-, thick, dashed, red] (u5) edge (u6);

\end{tikzpicture}  
        \caption{2-plex}
        \label{fig:model2}
    \end{subfigure}
    \hfill
    \begin{subfigure}{0.3\linewidth}
    \centering
        \newcommand{\x}{0.7}
\newcommand{\y}{0.6}

\newcommand{\scll}{1.1}
\newcommand{\scl}{1.25}

\newcommand{\xx}{5.8}
\newcommand{\yy}{-1}

\hspace{-9mm}

\begin{tikzpicture}  
  [scale=.6,auto=center,every node/.style={circle, inner sep=0.0mm, draw, minimum width=2mm, fill=gray}, > = stealth]

  \node [] (u1) at (-1*\x, 0*\x) {};
  \node [] (u2) at (1*\x, 0*\x) {};
  \node [] (u3) at (-2*\x, 1.732*\x) {};
  \node [] (u4) at (2*\x, 1.732*\x) {};
  \node [] (u5) at (-1*\x, 1.732*2*\x) {};
  \node [] (u6) at (1*\x, 1.732*2*\x) {};

  \path[-, thick, dashed, red] (u1) edge (u2); 
  \path[-, thick] (u1) edge (u3); 
  \path[-, thick] (u1) edge (u4); 
  \path[-, thick] (u1) edge (u5); 
  \path[-, thick] (u1) edge (u6); 
  \path[-, thick] (u2) edge (u3); 
  \path[-, thick] (u2) edge (u4); 
  \path[-, thick] (u2) edge (u5); 
  \path[-, thick] (u2) edge (u6); 
  \path[-, thick, dashed, red] (u3) edge (u4); 
  \path[-, thick] (u3) edge (u5); 
  \path[-, thick] (u3) edge (u6); 
  \path[-, thick] (u4) edge (u5); 
  \path[-, thick] (u4) edge (u6); 
  \path[-, thick, dashed, red] (u5) edge (u6);

\end{tikzpicture}  
        \caption{0.8-quasi-clique}
        \label{fig:model1}
    \end{subfigure}
    \hfill
    \begin{subfigure}{0.3\linewidth}
        \centering
        \newcommand{\x}{0.7}
\newcommand{\y}{0.6}

\newcommand{\scll}{1.1}
\newcommand{\scl}{1.25}

\newcommand{\xx}{5.8}
\newcommand{\yy}{-1}

\hspace{-9mm}

\begin{tikzpicture}  
  [scale=.6,auto=center,every node/.style={circle, inner sep=0.0mm, draw, minimum width=2mm, fill=gray}, > = stealth]

  \node [] (u1) at (-1*\x, 0*\x) {};
  \node [] (u2) at (1*\x, 0*\x) {};
  \node [] (u3) at (-2*\x, 1.732*\x) {};
  \node [] (u4) at (2*\x, 1.732*\x) {};
  \node [] (u5) at (-1*\x, 1.732*2*\x) {};
  \node [] (u6) at (1*\x, 1.732*2*\x) {};

  \path[-, thick] (u1) edge (u2); 
  \path[-, thick] (u1) edge (u3); 
  \path[-, thick] (u1) edge (u4); 
  \path[-, thick] (u1) edge (u5); 
  \path[-, thick] (u1) edge (u6); 
  \path[-, thick] (u2) edge (u3); 
  \path[-, thick] (u2) edge (u4); 
  \path[-, thick] (u2) edge (u5); 
  \path[-, thick] (u2) edge (u6); 
  \path[-, thick, dashed, red] (u3) edge (u4); 
  \path[-, thick] (u3) edge (u5); 
  \path[-, thick] (u3) edge (u6); 
  \path[-, thick] (u4) edge (u5); 
  \path[-, thick] (u4) edge (u6); 
  \path[-, thick] (u5) edge (u6);

\end{tikzpicture}  
        \caption{1-defective clique}
        \label{fig:model3}
    \end{subfigure}
    \caption{Relaxed cliques with six vertices (missing edges are represented by red dashed edges).}
    \label{fig:models}
\end{figure}

\begin{table}[t]
    \setlength{\tabcolsep}{3.85pt}
    \caption{Link prediction on PPI network, where \# cliq., \# mis., and Acc. denote the number of relaxed cliques, the number of missing edges, and the accuracy, respectively.}
    \centering
    \scalebox{0.85}{
    \hspace{-2mm}
    \begin{tabular}{c|ccc|ccc|ccc}
        \toprule
        \multirow{2}{*}{$k$} & \multicolumn{3}{c|}{$(1-0.1k)$-quasi-clique} & \multicolumn{3}{c|}{$(k+1)$-plex} & \multicolumn{3}{c}{$k$-defective clique} \\ 
        & \# cliq. & \# mis. & Acc. & \# cliq. & \# mis. & Acc. & \# cliq. & \# mis. & Acc. \\ \midrule
        $1$ & 143  & 64  & 82.8\% & 143  & 64  & 82.8\% & 56   & 40  & \textbf{92.5\%} \\ 
        $2$ & 476  & 113 & \textbf{88.5\%} & 508  & 113 & \textbf{88.5\%} & 324  & 69  & 82.6\% \\ 
        $3$ & 1020 & 279 & 74.2\% & 1405 & 279 & 74.2\% & 804  & 84  & \textbf{85.7\%}\\ 
        $4$ & 3521 & 671 & 44.9\% & 4831 & 671 & 44.9\% & 1482 & 108 & \textbf{85.2\%}\\ \bottomrule
    \end{tabular}
    \label{tab:ppi-predict}
    }
\end{table}

The $k$-defective clique has the following relationships with other relaxed cliques:
\begin{itemize}
\item A $k$-defective clique is also a $(k+1)$-plex.
\item A $k$-defective clique of size $q$ is also a $(1 - \frac{k}{q - 1})$-quasi-clique. 
\end{itemize}
For example, a 1-defective clique of size 6 in Figure~\ref{fig:model3} is also a $2$-plex  and a $1-\frac{1}{6-1}=0.8$ quasi-clique.
In our comparison, therefore, we evaluated $k$-defective cliques, $(k+1)$-plexes, and $(1 - \frac{k}{q - 1})$-quasi-cliques for various values of $k$.

We conducted an experiment to predict missing interactions in a protein–protein interaction (PPI) network~\cite{krogan2006global}, where each vertex represents a protein and each edge indicates an interaction. A protein complex is a group of proteins that are physically co-located within a cell and collectively perform a specific biological function. Protein complexes typically form cliques, i.e., every pair of proteins in the complex interacts with each other~\cite{yu2006predicting, bader2002analyzing}. Prior studies have attempted to identify missing interactions in noisy PPI networks by detecting relaxed cliques and completing the missing edges within them~\cite{yu2006predicting}. 
In our experiment, we evaluated the three relaxed clique models using the procedure below:

\begin{enumerate}[label=(\roman*)]
    \item Enumerate all maximal relaxed cliques of size greater than 10 (i.e., $q=11$).
    \item Identify all missing edges in detected relaxed cliques.
    \item Evaluate the accuracy, defined as the ratio of \emph{positive} missing edges, where a missing edge is considered positive~\cite{yu2006predicting} if the two proteins involved are members of the same protein complex according to a ground-truth dataset~\cite{pu2009up}.
\end{enumerate}

Table~\ref{tab:ppi-predict} shows the results. In most cases, the defective clique model achieves the highest prediction accuracy. Notably, 1-defective cliques attain the best performance, with an accuracy of 92.5\%. Moreover, as $k$ increases, the accuracy of $k$-defective cliques remains above 80\%, while the performance of other models drops significantly. This indicates that the defective clique provides a more accurate link prediction on PPI network compared to the other models.

\subsection{Parallelization}

\begin{figure}
    \centering

    \scalebox{\scaleval}{
          \begin{tikzpicture}
  \begin{axis}[
      hide axis,
      xmin=0, xmax=1,
      ymin=0, ymax=1,
      legend columns=6,
      legend style={draw=none, fill=none},
      height=2cm
  ]
  \addlegendimage{
  line width=0.3mm, color=\ParTTTTC, mark=\ParTTTTM, mark options={solid, scale=\ParTTTTSc}, every mark/.append style={rotate=\ParTTTTRo}
  }
  \addlegendentry{2 threads \quad\quad}
  
  \addlegendimage{
  line width=0.3mm, color=\ParTTTC, mark=\ParTTTM, mark options={solid, scale=\ParTTTSc}, every mark/.append style={rotate=\ParTTTRo}
  }
  \addlegendentry{4 threads \quad\quad}
  
  \addlegendimage{
  line width=0.3mm, color=\ParTTC, mark=\ParTTM, mark options={solid, scale=\ParTTSc}, every mark/.append style={rotate=\ParTTRo}
  }
  \addlegendentry{8 threads \quad\quad}
  
  \addlegendimage{
  line width=0.3mm, color=\ParTC, mark=\ParTM, mark options={solid, scale=\ParTSc}, every mark/.append style={rotate=\ParTRo}
  }
  \addlegendentry{16 threads}
  
  \end{axis}
  \end{tikzpicture}
    }

    \vspace*{-1.2mm}
    \hspace*{-4.5mm}
        \centering
        \scalebox{\scaleval}{\input{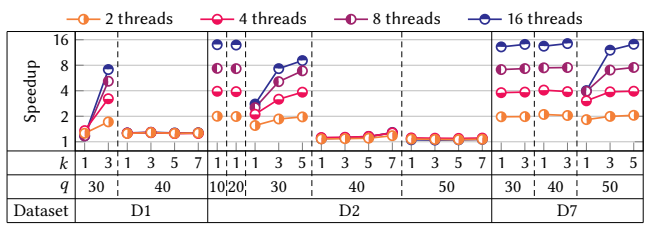}}

    \caption{Speedup achieved by the parallel implementation of \OursE{} compared to the sequential baseline.}
    \label{fig:maximal-parallel}
    \Description{}
\end{figure}

We implemented a parallel version of our enumeration algorithm \OursE{}. 
\OursE{} consists of two phases: graph reduction (Section~\ref{subsec:reduction}) and branch-and-bound (Algorithm~\ref{alg:diameter-two} in Section~\ref{subsec:twohop}). We have parallelized Algorithm~\ref{alg:diameter-two}.
Recall that Algorithm~\ref{alg:diameter-two} partitions the original task into $n$ subtasks, where each subtask enumerates all solutions whose smallest vertex is $v_i$ for each $v_i \in V$ (Lines~\ref{line3:V}--\ref{line3:branch}). Since these subtasks are independent, we dynamically assign them to multiple threads, allowing Lines~\ref{line3:foreach}--\ref{line3:branch} to be executed in parallel.

Figure~\ref{fig:maximal-parallel} shows the speedup achieved by the parallel algorithm using 2, 4, 8, and 16 threads, compared to the sequential algorithm (i.e., \OursE{}). 
In \emph{hard} cases, i.e., when the branch-and-bound phase is time-consuming (taking more than 60 seconds in Figure~\ref{fig:maximal-result}), our algorithm achieves nearly linear speedup with respect to the number of threads, e.g., on D2 when $q \le 20$ and on D7 when $q\le 40$. 
This is because, in hard cases, the problem is typically divided into a large number of subtasks, ranging from thousands to millions, which enables effective load balancing over multiple threads.
Cases where our algorithm did not exhibit linear speedup 
(e.g., on D1 and D2 when $q \ge 40$) 
correspond to \emph{easy} instances, where graph reduction accounted for the majority of the total processing time.

\color{black}

\section{Conclusion}
In this paper, we proposed a novel branch-and-bound algorithm for enumerating maximal $k$-defective cliques with a worst-case optimal search space. Based on this algorithm, we also developed an efficient framework for maximum $k$-defective clique search. We show the effectiveness of our algorithms both theoretically and empirically. 

It would be an interesting future work to apply our techniques to other problems related to relaxed cliques.
For example, our clique-first approach is a high-level idea that may be applied to other relaxed cliques.
Any relaxed clique consists of a large number of non-missing edges and a relatively smaller number of missing edges.
Our clique-first approach first generates a clique (i.e., non-missing edges) and then adds missing edges, which reduces the search space as shown in Figure~2. 
This approach may be applied to other relaxed cliques.

\balance
\bibliographystyle{ACM-Reference-Format}

\end{document}